\documentclass{appolb}

\pdfoutput=1

\usepackage{epsfig}
\usepackage{graphicx}
\usepackage{dcolumn}   
\usepackage{bm}        
\usepackage{amsmath}
\usepackage{amssymb}   
\usepackage{enumitem}
\usepackage{graphicx,mathrsfs}
\usepackage{nicefrac}
\usepackage{multirow}
\usepackage{color}

\usepackage{cmap}
\usepackage[utf8x]{inputenc}
\usepackage[greek,english]{babel}
\usepackage{textgreek}
\usepackage{ucs}
\usepackage{cite}

\begin{document}
\title{BFKL phenomenology%
\thanks{Presented at the Summer School and Workshop on High Energy Physics at 
the LHC: New trends in HEP, October 21- November 6 2014, 
Natal, Brazil}
}
\author{
G. Chachamis
\address{Instituto de F\'isica Te\'orica UAM/CSIC \& Universidad Aut\'onoma de
Madrid, C/ Nicol\'as Cabrera 15, E-28049 Madrid, Spain}
}
\maketitle
\begin{abstract}
We present some of the topics covered in a series of lectures under the same title that was given 
at the ``Summer School on High Energy Physics at the LHC: New trends in HEP" in Natal, Brazil. In 
particular, after some general thoughts on phenomenology we give a pedagogical introduction to 
the BFKL formalism and we discuss recent BFKL phenomenological studies for LHC observables.

\end{abstract}
\PACS{put PACS numbers here}
  
\def\as{\alpha_s}

\section{Introduction}

Phenomenology in its broader meaning, 
one would argue, has generally been instrumental 
in advancing the progress of human thought.
Despite the fact that the etymology of the term, from the Greek words 
{\it phain\'omenon} and {\it l\'ogos}, implies that phenomenology
is the study of `that which is observed', one has to stress that
there is not a unique definition for phenomenology,
or more accurately,  that the definition varies a lot
depending on the context (philosophy, psychology, science) and different
people from different origins may associate different notions with the term.

If we restrict ourselves in seeking a definition for phenomenology 
within the grounds of modern physics, the following directive 
provides a possible candidate:``Observe `that which appears', 
a collection of phenomena that share a unifying principle, and try to find patterns to describe it. 
The patterns might or might not be of fundamental nature or they might be up to a certain extend".
More specifically, for high energy physics, a possibly satisfactory statement could be the 
following: ``Use assumed fundamental laws to produce theoretical estimates 
for physical observables and then compare against experimental data to validate or falsify the assumed laws".

In the past decades, high energy physics was mostly studied in colliders
and the vast majority of experimentally measured quantities were observables stemming 
from the collision of particles. If the Standard Model (SM) enjoys such a wide
acceptance as the correct theory for the description of the Strong and the Electroweak (EW)
interactions, it has to do with a titanic effort from the experimental side 
(HERA, LEP,  Tevatron, LHC) and an equally important effort from the theory
community to provide theoretical estimates for a large amount of processes.
The comparison between theory and experiment results in favor of the SM and
so far no clear signal for new physics has emerged in any of the collision experiments.
It will be very interesting to see whether the second run of the LHC could change this picture.

Apart from the (generally rare) times that an experimental situation
is a standalone manifestation of a new phenomenon, 
it is usually after copious and demanding studies from both theoretical and
experimental sides that one can speak about agreement or disagreement 
of the predictions with the data. Focusing hereafter on the theory side, 
we could argue that SM phenomenology actually means computing estimates for
observables by employing
perturbation theory since the SM Lagrangian cannot be solved exactly.
We know that perturbation theory is only an approximation and
cannot be applied without the presence of a small expansion parameter.
The usual small parameter is the EW coupling in calculations in the EW
sector of the SM and the strong coupling $\as$ in QCD. Moreover, in hadronic colliders,
practically no physical observable lives solely 
 in a region of the phase space\footnote
{The term ``phase space" here is to be understood as a very
wide notion: all possible configurations of initial conditions connected to all possible configurations
of final states consist the phase space of observables.} 
where  non-perturbative input is unnecessary. This becomes evident
for LHC observables if we think that the colliding particles are protons,
objects of a non-perturbative nature due to their size and structure. Various
factorization theorems and schemes are employed to put
 some order to that picture. The main idea behind
factorization is that one separate the hard (perturbative) from the soft (non-perturbative)
physics  such that in order to have a theoretical estimate for an observable one needs to 
calculate the contribution from hard physics to that process
 and convolute it with a parametrization
of the soft physics involved that takes into account all available data. The parametrization 
is based on the fact that soft physics can be in general process independent, e.g.
the proton PDF's describe the probability to find  a certain parton in the proton
disregarding of the process in which the proton is involved.

The main bulk of phenomenological studies in the past decades is based
on the so-called `fixed order' calculations in which
one considers only the first few terms 
of the perturbative expansion for a  (hard) process  and computes these
terms fully. The perturbative expansion is realized via  Feynman calculus
and each term is graphically represented by Feynman diagrams.
Assuming only the first term results to leading order (LO) calculations, assuming
the first and second term results to next-to-leading (NLO) calculations, assuming the
first three terms leads to next-to-next-to leading (NNLO) calculations and so forth.
Most of the LHC processes require theoretical prediction at least to NLO 
and some of them to NNLO accuracy for a definite answer after comparing against
experimental data. The complexity increases enormously as one goes from one
order to the next and also as the number of external particles that participate in
the process increases. Fixed order calculations justify fully their reputation
of being `precision physics' calculations since the only uncertainty that remains at the end
is the uncertainty from omitting the higher term contributions and this can
be well estimated in most cases\footnote{There is also uncertainty
from the non-perturbative input (e.g. PDF's) but this is not directly
connected to the fixed order calculation of the hard part of a process,
or at least this is what factorization dictates.}.

In many cases, especially in hadron collider processes,
not only a fixed order calculation is too complicated
to be done beyond LO (e.g. multi-jet production) but also we have the
presence of a large scale (usually the logarithm of a
kinematical invariant or a mass) that persistently appears in every
order combined with the small expansion parameter 
in a certain way and potentially could break
down the convergence of the perturbative expansion. In such cases,
we need a resummation scheme to sum all the large contributions
from all terms (to order infinity). The result of the resummation can
either be combined with a fixed order calculation or, if it encodes truly
the most important contributions of each term in the expansion, it can be used
alone as the theoretical estimate for a hard process. We should stress
that any resummation approach is to be understood in the context of
perturbation theory and also that each term of the
perturbative expansion is represented by (effective) Feynman diagrams.

One of the most important resummation programs appears in the context of
high energy scattering. 
If in a process the center-of-mass energy, $\sqrt{s}$, is really large 
then the product $(\as \ln s)$ can easily be of order unity. If in addition $s$
is much larger than any other scale present, then in principle 
any term $\sim (\as \ln s)^n$, where $n$ is arbitrarily high, would give
the main contribution of the $n$-th term of the expansion and
this term cannot be omitted. Instead, one has to resum all these
important contributions up to $n \rightarrow \infty$. This is done within the
Balitsky-Fadin-Kuraev-Lipatov (BFKL) formalism at leading 
logarithmic (LA)~\cite{Lipatov:1976zz,Fadin:1975cb,Kuraev:1976ge,Balitsky:1978ic}
and at next-to-leading logarithmic (NLA) accuracy~\cite{Fadin:1998py,Ciafaloni:1998gs}.
For the latter, also terms that behave like $\as (\as \ln s)^n$  are resummed.
One sees thus, that  resummation programs can also be regarded as a new
perturbative expansion: the first term contains all the leading logarithmic terms
to all orders (LO approximation), 
the second term the sub-leading logarithms (NLO corrections) and so on.

In the next section we will sketch a derivation of the BFKL equation that 
resume all large logarithms in $s$ after introducing some important notions
that are ubiquitous in the BFKL framework and of which the origin or 
the relevance are not obvious to the non-expert.
In Section 3 we will discuss recent phenomenological studies for
BFKL related observables and in Section 4 we will conclude with
a general discussion.

\section{The BFKL equation and the Pomeron}

\begin{figure}
\centerline{\epsfig{file=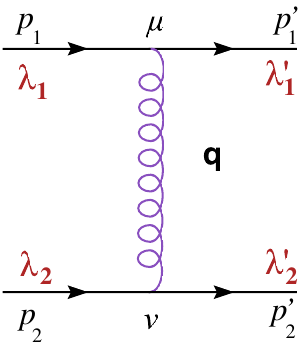,height=6cm}}
\caption{$qq$-scattering at LO order.}
\label{fig:nlip00}
\end{figure}

In this section, using a diagrammatic
approach, our aim is to see

\begin{itemize}
\item How logarithms in $s$ make
their appearance in high energy scattering
\item That these logarithms appear in all orders of the perturbative expansion
\item How to resum these logarithms.
\end{itemize}

Setting our goals as listed above accounts as
a minimal but hopefully an honest try to gain a first
insight in BFKL physics. Following this course of reasoning though,
will actually permit us to 
see a lot more, albeit on a very pedagogical level only. The ambitious reader
who wants a deeper insight should consult more complete
presentations of the topic, for example the works in
Refs.~\cite{nbarone, nforshawross,nqcdlev,DelDuca:1995hf,Salam:1999cn,Fadin:1998sh}
and the original publications which arguably hide a richness of
thought that cannot be covered in any review article.

Nevertheless, even in this minimalistic setting of goals as presented above
and while we are chasing $(\as \ln s)^n$-like
terms in Feynman diagrams we will still be able to see
\begin{itemize}
\item How we  separate virtual from real corrections and treat them 
in a separate manner and why this is of great importance.
\item What the reggeization of the gluon is and pinpoint its origin.
\item The different role of longitudinal and transverse degrees of freedom
\item The ``derivation" of the BFKL equation and its solution for the forward case.
\item What the Pomeron is and whether we can describe it in a simple manner.
\end{itemize}

Let us start by considering $q q$-scattering at lowest order (Born level) as 
depicted in Fig.~\ref{fig:nlip00}. Our discussion will be based a lot on the
way the topic is presented in  \cite{nforshawross, nbarone}.
Since we are concerned with high energy scattering, we will
work in the high energy limit which is defined by the condition
\begin{equation}
s \gg |t| ,\, u \simeq -s\,.
\label{eq:nreggel}
\end{equation}
The two quarks are interacting
via a gluon exchange in the $t$-channel. 
We can write the momentum of the gluon in
Sudakov parameters:
\begin{equation} 
q\,=\,\rho \,p_1\,+\,\sigma\, p_2\,+\,  q_{\perp}\,,
\end{equation}
where $p_1$ and $p_2$ are the momenta of the incoming quarks and
$q_{\perp} = (0, {\bf q_{\perp}}, 0)$ is a four-vector with non-zero entries for
only the transverse part of the gluon momentum. 
To denote two-dimensional transverse vectors we use boldface characters hereafter.
To keep contact with the physical picture of a collision
in an experiment, any transverse momentum in the following should be
understood as the projection of the total momentum on the transverse 
to the beam axis plane.
Our kinematical invariants then expressed in Sudakov variables read:
$s\,=\,2 p_1 p_2$ and $t = q^2 = \rho~\sigma s - {\bf q}^2$. 
We should keep in mind that 
for perturbation theory to apply, we need the presence of a hard
scale $Q$ that will ensure the smallness of the strong coupling $\alpha_s(Q)$.
We assume that such a scale  exists but we leave it unidentified for the moment. 
Moreover, all  factors in the formulae to follow
that are irrelevant to the kinematics (such as color factors)
will be suppressed. 

For the upper vertex  in Fig.~\ref{fig:nlip00} we have:
\begin{equation}
-i g_s  \bar{u}( p_1+q ) \gamma_{ \mu } u( p_1)\,.
\end{equation}
Because of Eq. \ref{eq:nreggel}, $q \ll p_1$ and
the above formula can be approximated by
\begin{equation}
-i g_s \bar{u}(p_1) \gamma_{\mu} u(p_1) \simeq -2 i g_s p_1^{\mu}\,.
\label{eq:nikon}
\end{equation}
Approximating similarly the lower vertex, the amplitude for the
process at hand at LO reads
\begin{equation}
A^{(0)}(s,t)=8 \pi a_s ~\mathcal{CF}_1 \frac{s}{q^2}=8 
\pi a_s ~\mathcal{CF}_1 \frac{s}{t}\,.
\label{eq:ntreeam}
\end{equation}
where $\mathcal{CF}_1$ denotes a color factor.
We see that there are no logarithms in $s$
in Eq.~\ref{eq:ntreeam} as was easy to guess beforehand. We would like
now to move to the next order and consider diagrams that stem from
the tree diagram after attaching another gluon. This new gluon can
either be virtual, in which case we will have one-loop diagrams (virtual
radiative corrections), or it can be real which would mean that it could in
principle be detected in the final state (real corrections).
Instead of considering both real and virtual corrections simultaneously 
at NLO, we will follow a different course. We will consider first only
the virtual correction, first at NLO, then at NLLO and see where this
approach can take us.

\begin{figure}
\centerline{\epsfig{file=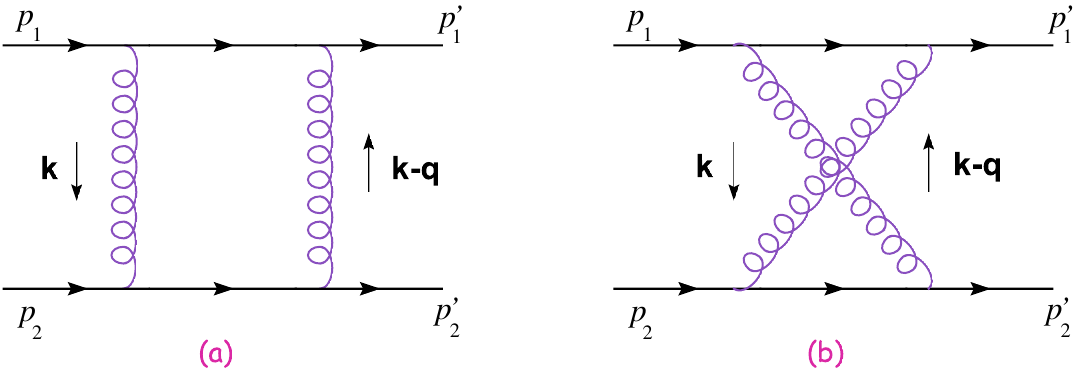,height=4.5cm}}
\caption{$qq$-scattering, one-loop corrections.}
\label{fig:nlip11}
\end{figure}

\begin{figure}
\centerline{\epsfig{file=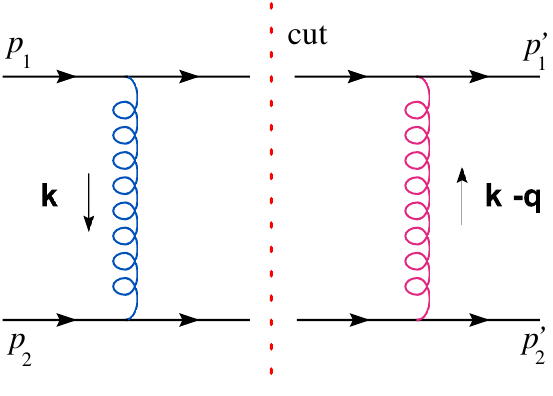,height=5cm}}
\caption{$qq$-scattering, one-loop cut amplitude.}
\label{fig:nlip11_}
\end{figure}

It turns out that
one-loop diagrams with self-energy and vertex corrections
are sub-leading in $\ln s$ and do not need to be computed.
Only box diagrams are contributing and the ones
 that give the relevant $\ln s$ term are shown in
Fig.~\ref{fig:nlip11}.
Let us focus on the Fig.~\ref{fig:nlip11}(a) diagram.
We can calculate its imaginary part using the  
Cutkosky rules (see Fig~\ref{fig:nlip11_}) and then obtain the full amplitude 
by dispersion relations.
Denoting the NLO amplitude by $A^{(1)}(s,t)$ we have:
\begin{equation}
Im A^{(1)}(s,t)=\frac{1}{2} \int d \mathrm{PS}^{(2)} A^{(0)}(s,k^2)  A^{(0)\dagger}(s,(k-q)^2)\,,
\label{eq:n1loop}
\end{equation}
where $A^{(0)}(s,k^2)$ and $A^{(0)\dagger}(s,(k-q)^2)$ are the tree level amplitudes
in Fig.~\ref{fig:nlip11_} with the quark lines being on shell at the cut points.
$A^{(0)\dagger}$ stands for the hermitian conjugate of
$A^{(0)}$.
The two-body phase space $\int d \mathrm{PS}^{(2)}$ is given by
\begin{equation}
\int d \mathrm{PS}^{(2)} = \int \frac{d^4 k}{(2\pi)^2} \delta((p_1-k)^2) \delta((p_2+k)^2)\,.
\end{equation}
Again, by introducing Sudakov variables $\rho$, $\sigma$ 
we can express  $k$ and $d^4 k$ as
\begin{equation}
k = \rho \,p_1 + \sigma\, p_2 + k_{\perp}, \,\,\,\, d^4 k = \frac{s}{2}d \rho d \sigma d^2 {\bf k}\,.
\end{equation}
so that we finally obtain for the two-body phase space:
\begin{equation}
\int d \mathrm{PS}^{(2)}  =\frac{1}{8 \pi^2 s} \int d^2 {\bf k}\,.
\label{eq:nfacespa}
\end{equation}
The two tree level amplitudes in Eq.~\ref{eq:n1loop} read
\begin{equation}
A^{(0)}(s,k^2)\,=\,-8 \pi a_s \mathcal{CF}_2 \frac{s}{{\bf k}^2}
\end{equation}
and
\begin{equation}
A^{(0)\dagger}(s,(k-q)^2)\,=\,-8 \pi a_s \mathcal{CF}_3
\frac{s}{({\bf k} - {\bf q})^2},
\end{equation}
where $\mathcal{CF}_2$ and $\mathcal{CF}_3$ are color factors.
The imaginary part of $A^{(0)}(s,t)$, with the help of Eq.~\ref{eq:nfacespa}, becomes:
\begin{equation}
\mathrm{Im} A^{(1)}(s,t)\,=\,4  \alpha_s^2\,  s\,\mathcal{CF}_4
 \int \frac{d^2 {\bf k}}{{\bf k}^2 ({\bf k}-{\bf q})^2}\,
\end{equation}
and by dispersion relations we can reconstruct the full amplitude which reads:
\begin{equation}
A^{(1)}(s,t)\,=\, -4 \frac{ \alpha_s^2}{\pi}\, \mathcal{CF}_4 \ln(\frac{s}{t}) \,s 
\int \frac{d^2 {\bf k}}{{\bf k}^2 ({\bf k}-{\bf q})^2}\,.
\label{eq:nfuloop}
\end{equation}
We remind the reader that we are tracing leading logarithms in $s$, and since $s/t<0$
we can write for a generic amplitude $\mathcal{A}\sim
\mathcal{B} \ln \frac{s}{t}$ after making the decomposition into real and imaginary parts:
\begin{equation}
\mathcal{A}=\mathrm{Re} \mathcal{A}+ i ~\mathrm{Im} \mathcal{A} \sim
\mathcal{B} \ln \frac{s}{t} = \mathcal{B} \ln \frac{s}{|t|}- i \pi \mathcal{B}
\end{equation}
which simply means
$\mathrm{Re} \mathcal{A}\,=\,-\frac{1}{\pi}\, Im \mathcal{A} \,\ln \frac{s}{|t|}$\,.
Thus,  after defining 
\begin{equation}
\epsilon(t) = \frac{N_c  \alpha_s}{4 \pi^2}\, \int\,- { \bf q}^2 \,
\frac{ d^2 {\bf  k}}{{\bf k}^2({\bf k}-{ \bf q})^2}\,,
\label{eq:ngluontrajectoraki}
\end{equation}
where the function $\epsilon(t)$ is called gluon Regge trajectory, we  rewrite 
Eq. \ref{eq:nfuloop} as 
\begin{equation}
A^{(1)}(s,t) \,=\,-\frac{16 \pi \alpha_s}{N_c}\, \mathcal{CF}_4 \frac{s}{t} \,\ln(\frac{s}{t})\,\epsilon(t)\,,
\label{eq:nfuloop3}
\end{equation}
whereas for the Fig.~\ref{fig:nlip11}(b) diagram in the crossed channel it will be:
\begin{equation}
A_{\mathrm{cross}}^{(1)}(s,t) \,=\,-\frac{16 \pi \alpha_s}{N_c}\,\mathcal{CF}_5 
 \frac{u}{t} \,\ln(\frac{u}{t})\,\epsilon(t)\,.
\label{eq:nfuloop3u}
\end{equation}
\begin{figure}
\centerline{\epsfig{file=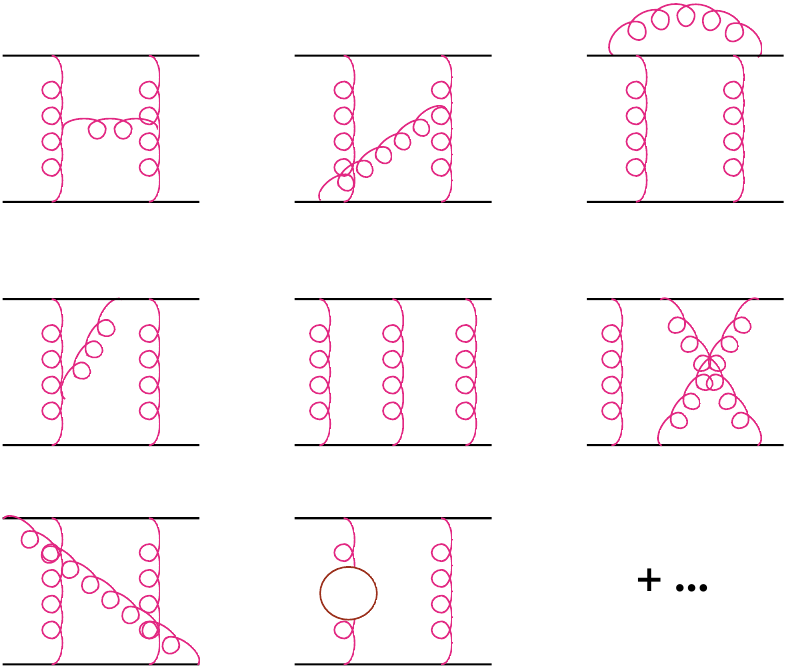}}
\caption{$qq$-scattering, two-loop diagrams.}
\label{fig:nlipmany1}
\end{figure}
After adding the last two equations and using $u \simeq -s$,
we obtain the one-loop amplitude. Considering
colour octet exchange\footnote{We have hidden any color dependence
of the amplitudes in the
color factors $\mathcal{CF}_i$, any reader interested in color decomposition 
should consult Ref.~\cite{nqcdlev},
in particular, Section 9.4.3.}, 
we can express the one-loop amplitude in terms of the tree level one, specifically:
\begin{equation}
A_8^{(1)}(s,t) = 8 \pi a_s \mathcal{CF}_1 ~ \frac{s}{t}
\,\ln(\frac{ s }{|t|})\,\epsilon(t)\, =
\,A^{(0)}\,\ln( \frac{s}{|t|})\,\epsilon(t)\,.
\label{eq:noctet1}
\end{equation}

\begin{figure}
\centerline{\epsfig{file=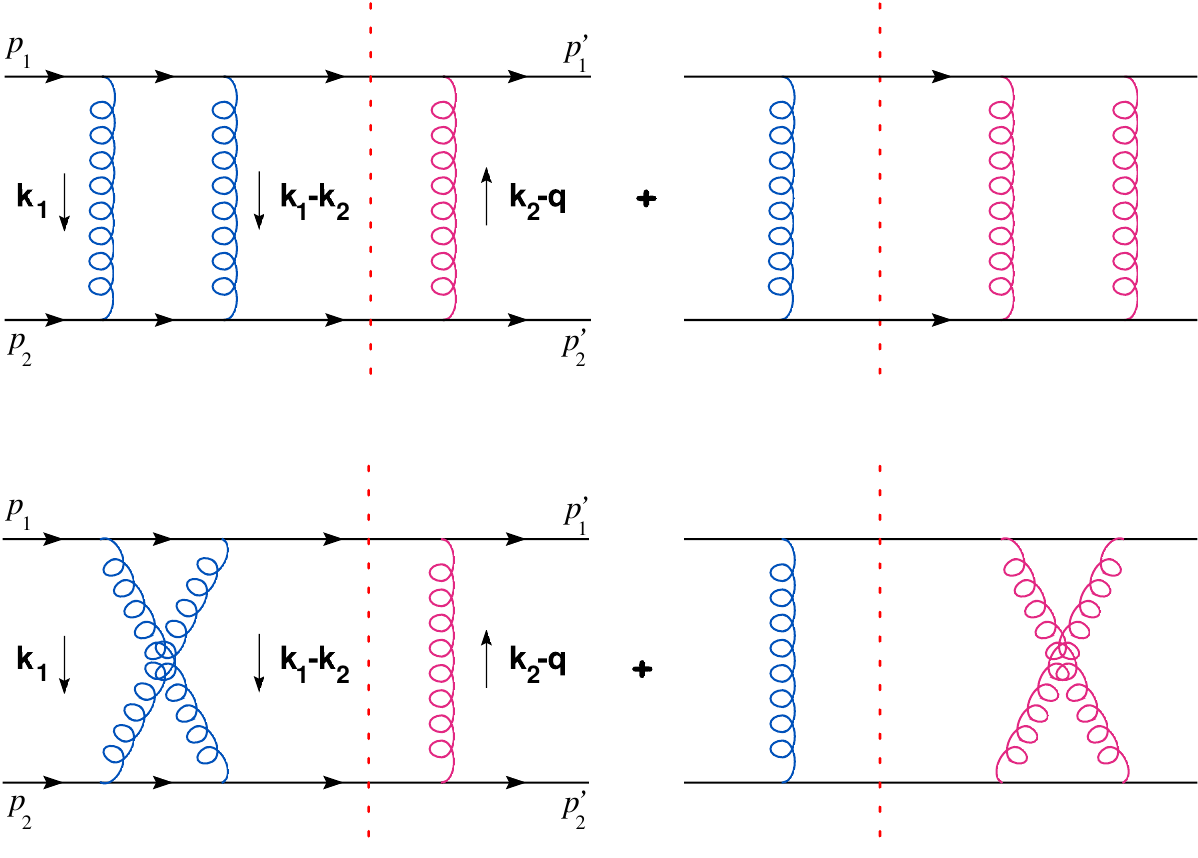}}
\caption{$qq$-scattering, two-loop box virtual corrections.}
\label{fig:nlipvirt1}
\end{figure}

One order higher in the perturbative expansion, to $\mathcal{O}(\alpha_s^3)$,
we have to consider many Feynman diagrams like the ones in Fig.~\ref{fig:nlipmany1}.
Not all of them though contribute with leading logarithms.
The ones we need to compute are box diagrams, in particular,
 the two-loop box diagrams in Fig.~\ref{fig:nlipvirt1}.
Using the Cutkosky rules again,
we can express the two-loop diagrams into
one-loop and tree contributions that are known
from the analysis so far. Indeed, in Fig~\ref{fig:nlipvirt1},
after multiplying the amplitudes to the left of the cut line 
by the (hermitian conjugates of) the ones to the right, summing over helicities
and performing the integration over the phase space,
we reach the very interesting result:
\begin{equation}
A_8^{(2)}(s,t) = \, A^{(0)}(s,t) \,\frac{1}{2} \ln^2(\frac{s}{|t|})\,\epsilon^2(t)\,,
\label{eq:noctet2}
\end{equation}
where the two-loop amplitude is expressed in terms of the LO one.
The expressions for  $A_8^{(2)}(s,t)$ and $A_8^{(1)}(s,t)$  
tell us that the partial result for the amplitude up to order  $\mathcal{O}(\alpha_s^3)$ is
\begin{equation}
A^{\mathrm{partial}}_8(s,t) = \, A^{(0)}(s,t) \,\left( 1 + \ln(\frac{s}{|t|})\,\epsilon(t) + \frac{1}{2} \ln^2(\frac{s}{|t|})\,\epsilon^2(t) \right)\,.
\label{eq:npartial}
\end{equation}
suggesting that the all-orders virtual amplitude might be of the form
\begin{equation}
A_8(s, t) = \, A^{( 0 )}(s, t) \,\left( 1 + \ln(\frac{ s}{|t|})\, \epsilon(t) 
+ \frac{1}{2} \ln^2 ( \frac{s}{|t|})\,\epsilon^2( t )
+ . . . \right)\,,
\label{eq:noctetall}
\end{equation}
namely, a product of the tree level amplitude and something that looks very much
like a series expansion. From that point on, it only takes a small logical step
to postulate that
\begin{equation}
A_8(s,t) = \, A^{(0)}(s,t) \,\left(\frac{s}{|t|}\right)^{\epsilon(t)}.
\label{eq:nexpansion}
\end{equation}
It is impressive to know that the ansatz  
in Eq.~\ref{eq:nexpansion} is proven to be true by the so-called  bootstrap equation.

At this point, we have partially achieved 
one of our primary goals, we have seen how logarithms in $s$
appear in virtual diagrams in different orders of the perturbative expansion and we
have managed to resum them in a closed form to all orders. The final result can
be written in a factorized form involving two terms, the Born amplitude and the expression
$\left(\frac{s}{|t|}\right)^{\epsilon(t)}$ which accounts for the resummation of
the large energy logarithms. We can actually obtain Eq.~\ref{eq:nexpansion}
by going back to Fig.~\ref{fig:nlip00} and calculating the tree level amplitude using for
the $t$-channel gluon a modified propagator which would  read:
\begin{equation}
D_{\mu \nu}(s,q^2)=-i \frac{g_{\mu \nu}}{q^2}\left(\frac{s}{{\bf k}^2}\right)^{\epsilon(q^2)}\,.
\label{eq:nregluon}
\end{equation}
Eq.~\ref{eq:nregluon} states that in the high energy limit, in order to take
into account all the important contributions from virtual diagrams to all orders
it suffices to calculate the tree level amplitude using a modified
propagator for the $t$-channel gluon. The importance
of this striking result cannot be overestimated.
The gluon with the modified propagator is called a reggeized gluon
or Reggeon and it hints that  the relevant degrees
of freedom in high energy scattering might not be 
just quarks and gluons.

\begin{figure}
\centerline{\epsfig{file=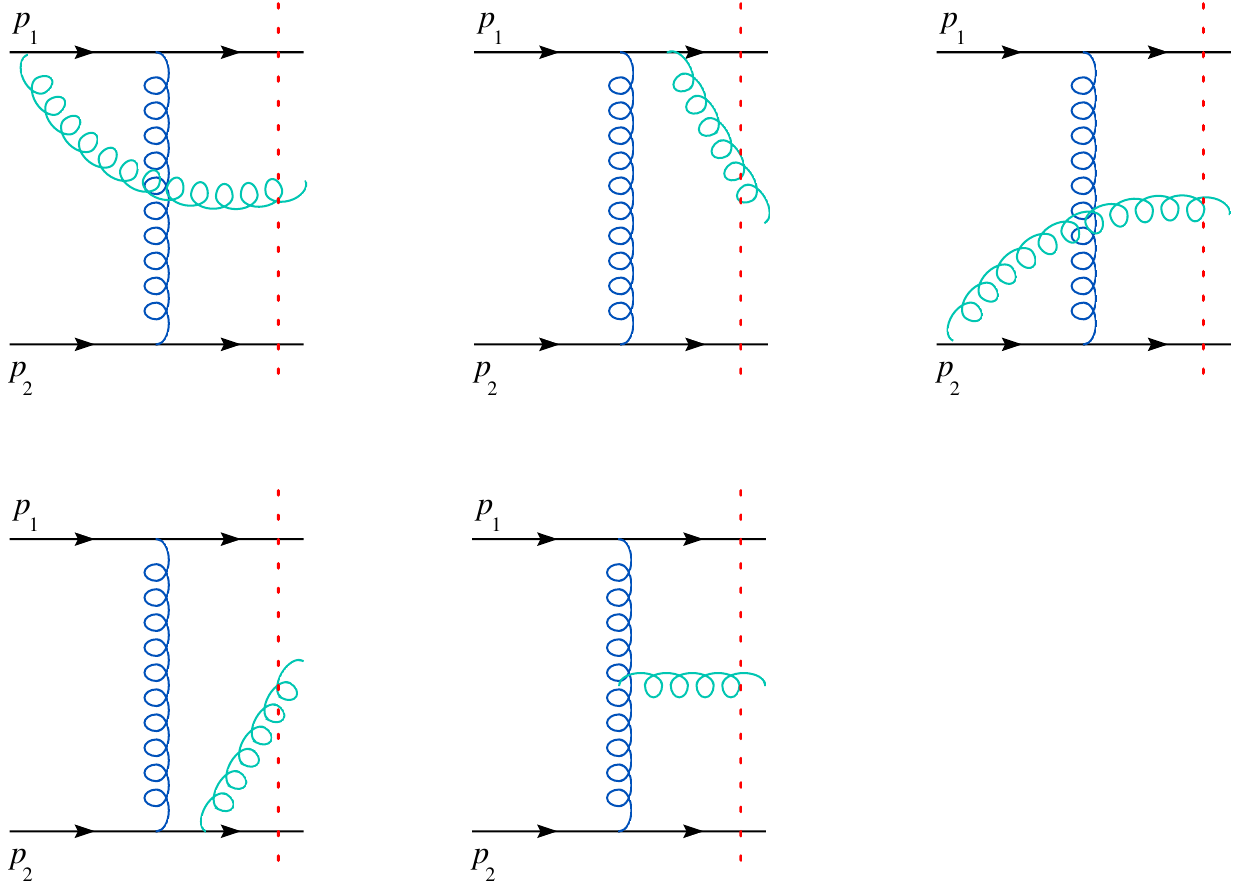,height=7cm,width=10cm}}
\caption{$qq$-scattering, one real gluon emission.}
\label{fig:nlipreal1}
\end{figure}

Let us now focus on the real corrections and in particular 
the real gluon emission
diagrams in Fig.~\ref{fig:nlipreal1} which are the first real emission
corrections to the Born amplitude. Formally these are 
 $\mathcal{O}(\alpha_s^3)$ corrections.

\begin{figure}
\centerline{\epsfig{file=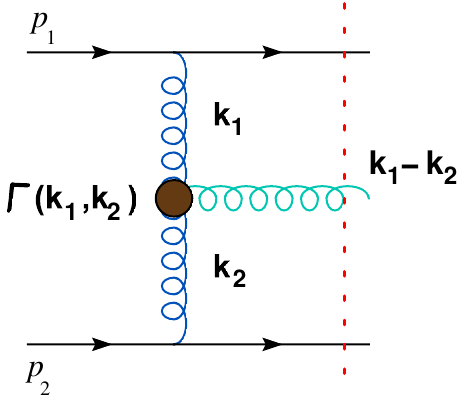,height=6cm}}
\caption{The Lipatov effective vertex.}
\label{fig:nlipeff1}
\end{figure}

It turns out that instead of calculating the amplitudes for all 
those diagrams it suffices to substitute their contribution
by the diagram in Fig.~\ref{fig:nlipeff1}
where the blob stands for the  Lipatov effective vertex
which is gauge invariant and has a tensorial structure.
The Lipatov effective vertex sums
the contributions from the graphs in Fig.~\ref{fig:nlipreal1}
 in an elegant way.
Using once more
Sudakov decomposition, the momenta of the two $t$-channel gluons 
in Fig.~\ref{fig:nlipeff1} read
\begin{eqnarray}
k_1&=&\rho_1 p_1 + \sigma_1 p_2 + {k_1}_{\perp}\nonumber\\
k_2&=&\rho_2 p_1 + \sigma_2 p_2 + {k_2}_{\perp}\,,
\label{eq:nsuda}
\end{eqnarray}
 and the relevant kinematical limit is given by the following conditions:
\begin{eqnarray}
1\,&>>\,\rho_1&>>\,\rho_2\nonumber\\
1\,&>>\,|\sigma_2|\,&>>|\sigma_1|
\label{eq:norder}
\end{eqnarray}
Using the Cutkosky rules once more,
we contract the tree level amplitude from
the diagram in Fig.~\ref{fig:nlipeff1} with its hermitian conjugate
and we integrate over the three-body phase space which
in our Sudakov parametrization reads
\begin{eqnarray}
\int d \mathrm{PS}^{(3)}\,=\,\frac{s^2}{4 (2\pi)^5}\, \int d \rho_1 d\rho_2 d\sigma_1 d\sigma_2
d^2{\bf k}_1 d^2{\bf k}_2\nonumber\\
\delta(-\sigma_1(1-\rho_1)s-{\bf k}^2_1)\,\,
\delta(\rho_2(1+\sigma_2)s-{\bf k}^2_2)\nonumber\\
\delta((\rho_1-\rho_2)(\sigma_1-\sigma_2)s-({\bf k}_1-{\bf k}_2)^2)\,.
\label{eq:n3body1}
\end{eqnarray}
Because of Eq. \ref{eq:norder} we may use the following approximations:
\begin{eqnarray}
1-\rho_1\,&\simeq& 1,\nonumber\\
1+\sigma_2\,&\simeq& 1,\nonumber\\
\rho_1-\rho_2\,&\simeq& \rho_1,\,\,\,\sigma_1-\sigma_2\,\simeq -\sigma_2\,,
\end{eqnarray}
so that Eq.~\ref{eq:n3body1} now reads
\begin{eqnarray}
\int d \mathrm{PS}^{(3)}\,=\,\frac{s^2}{4 (2\pi)^5}\, \int d \rho_1 d\rho_2 d\sigma_1 d\sigma_2
d^2{\bf k}_1 d^2{\bf k}_2\nonumber\\
\delta(-\sigma_1 s-{\bf k}^2_1)\,\,\delta(\rho_2 s-{\bf k}^2_2)\,\,
\delta(-\rho_1 \sigma_2 s-({\bf k}_1-{\bf k}_2)^2)\,.
\label{eq:n3body2}
\end{eqnarray}
It is from the rightmost delta function in (Eq.~\ref{eq:n3body2})
that the $\ln s$ behavior of the real corrections arises. Indeed,
after carrying out the integration over $\sigma_2$, 
the remaining integrand will acquire 
an $(1/\rho_1)$ factor:
\begin{eqnarray}
\int d \mathrm{PS}^{(3)}\,=\,\frac{1}{4 (2\pi)^5 s}\, \int^1_{{\bf k}_2^2/s}\,\frac{d \rho_1}{\rho_1}
\,\int  d^2{ \bf k}_1 d^2{ \bf k}_2
\label{eq:n3body3}
\end{eqnarray}
and finally performing the $\rho_1$ integration yields a factor
\begin{equation}
\ln \left(\frac{s}{{\bf k}_2^2}\right)\,=\,\ln \left(\frac{s}{ s_0}\right)\,,
\end{equation}
where  $s_0$ is a typical momentum, a
typical normalisation scale.

To consider one order higher corrections, we need to consider two real gluon emissions.
The diagrammatic depiction would be the one in Fig~\ref{fig:nlipeff1}
but now with two Lipatov effective vertices and three gluon propagators in
the $t$-channel. We would need to integrate over the four-body phase space 
in order to get the leading logarithms in $s$. It is straightforward 
to generalise this procedure for three, four and finally an arbitrary
number of real gluon emissions. 
We would like at this point to find a way to
combine the real with the virtual corrections
and most importantly, to find a way to account for the real emission corrections
to all orders in an closed form expression.

\begin{figure}
\centerline{\epsfig{file=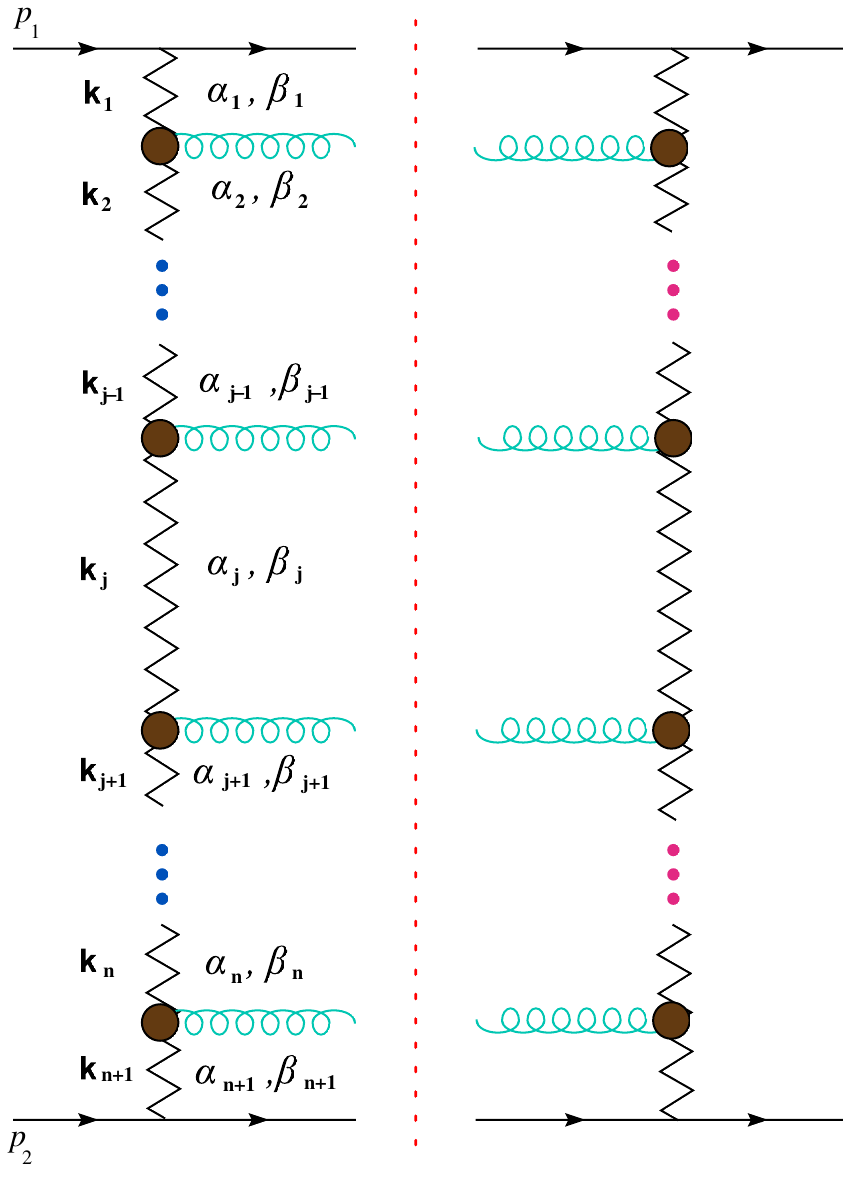,height=14cm}}
\caption{A typical gluonic ladder diagram.}
\label{fig:nlipladder1}
\end{figure}

Let us recapitulate here what insight we have gained and
assess where where stand with regard to our initial aims. In the 
discussion about the virtual corrections,
we have introduced the notion of  gluon reggeization:
a $t$-channel gluon with a modified propagator defined as in Eq.~\ref{eq:nregluon}
takes into account the leading logarithmic 
contributions from virtual diagrams to all orders. This is the closest one
can have for a recipe: to account for virtual corrections, substitute
the $t$-channel gluon by a Reggeon. On the other hand, the
idea of combining the various one real emission diagrams into 
a single diagram where we consider
one gluon emission in the $s$-channel that connects
to the $t$-channel gluon by means of a  Lipatov effective vertex
allows for the iteration of this prescription to cover
an arbitrary high $n$-gluon emissions.
All these lead very naturally to what we call  ladder diagrams,
an example is depicted in Fig.~\ref{fig:nlipladder1}.
This is the general picture of a BFKL ladder
in the colour singlet exchange, 
and a graphical depiction of what we call the perturbative Pomeron.
Let us have a closer view at the diagram in Fig \ref{fig:nlipladder1}. It consists of
$n$ rungs (real emitted gluons) connected to the $t$-channel reggeized gluons
(zig-zag lines) via
Lipatov effective vertices. The $t$-channel gluons 
are partitioned into $ n+1 $ reggeized propagators. 
The imaginary part of the amplitude, $\mathrm{Im} \mathcal{A}(s,t)$
for a process like that will be given
by convoluting the two tree level amplitudes (left and right to the cut) and
after integrating over the $n+2$-body phase space. 
The generalisation of the condition in Eq.~\ref{eq:nreggel}, 
leads to the kinematical configuration called 
multi-Regge kinematics (MRK):
\begin{eqnarray}
\hspace{-1.0cm}{\bf k}^2_1 \,\simeq\, {\bf k}^2_2\,\simeq\,...\, {\bf k}^2_i\,\simeq\, 
{\bf k}^2_{ i+1}\,...\,\simeq\, {\bf k}^2_{n}\,\simeq\, {\bf k}^2_{ n+1}\,\gg\,{\bf q}^2\,\simeq\,s_0,\nonumber\\
\hspace{-3cm}1\,\gg\,\rho_1\,\gg\,\rho_2\,\gg\,...\,\rho_i\,\gg\,\rho_{i+1}\,\gg
\,\rho_{n+1}\,\gg\,\frac{s_0}{s},\nonumber\\
\hspace{-3cm}1\,\gg\,|\sigma_{n+1}|\,\gg\,|\sigma_n|\,\gg\,...\,\gg\,|\sigma_2|\,\gg\,|\sigma_1|\,\gg\,\frac{s_0}{s}\,.
\label{eq:nMRK}
\end{eqnarray}
The nested integration over the phase space is a nuisance and the way out is
 to turn the 
multi-nested integral into a product of integrals by taking the Mellin transform 
of $\mathrm{Im} \mathcal{A}(s,t)$ working thus,  in the complex
angular momentum space $\omega$:
\begin{eqnarray}
f(\omega,t)\, = \,\int_1^{\infty} d\left(\frac{s}{s_0}\right)\,\left(\frac{s}{s_0}\right)^{-\omega-1}
\frac{\mathrm{Im} \mathcal{A}(s,t)}{s}\,.
\label{eq:nmell}
\end{eqnarray}
$ f(\omega,t) $ can be the staring point from which
we further define a function $f_{\omega}({\bf k}_a,{\bf k}_b,t)$ which, as its arguments
suggest, is the Mellin transform of the amplitude with the integrations
over the transverse momenta ${\bf k}_a$ and
${\bf k}_b$  still to be performed, where ${\bf k}_a$ and ${\bf k}_b$ are the topmost and bottommost
reggeized gluon propagators in the ladder. The function 
$f_{\omega}({\bf k}_a,{\bf k}_b,t)$ is the so-called BFKL Green's function.
Since $t \simeq -{\bf q}^2$, we will prefer the notation $f_{\omega}({\bf k}_a,{\bf k}_b,{\bf q}^2)$
in the following, in which the propagators ${\bf k}_a$ and $(q - {\bf k}_b)^2$
are contained and with ${\bf q}^2$ we denote the momentum transfer
in the $t$-channel. One could then
take $n=1$ in the ladder diagram in Fig.~\ref{fig:nlipladder1} and calculate the
corresponding $f^{(1)}_{\omega}({\bf k}_a,{\bf k}_b,{\bf q}^2)$ function
and then set $n=2$ and calculate the $f^{(2)}_{\omega}({\bf k}_a,{\bf k}_b,{\bf q})$
and after iterating this procedure up to an arbitrary $n \rightarrow \infty$ and summing up
all contributions,
one would compute $f_{\omega}({\bf k}_a,{\bf k}_b,{\bf q}^2)$.
Easy to describe but impossible to do. Instead, there is an elegant way through.
After taking the Mellin transform in Eq.~\ref{eq:nmell} and writing the generic expression
for $f_{\omega}({\bf k}_a,{\bf k}_b,{\bf q}^2)$ with the phase space integration still
to be done, one realizes\footnote{To demonstrate that, one needs to go through the
calculation which is beyond our scope here. The reader is encouraged
to try it out with the help of the references cited in the beginning of the section
in order to see how magic works.}
that there exists an integral equation
which governs the behavior of $f_{\omega}$:
\begin{eqnarray}
\omega f_{\omega}({\bf k}_a,{\bf k}_b,{\bf q})=\delta^2({\bf k}_a-{\bf k}_b)\nonumber\\
+ \frac{\bar{\alpha_s}}{2  \pi} \int d^2{\bf l} 
\bigg\{
\frac{-{\bf q}^2} {({\bf l}-{\bf q})^2 {\bf k}_a^2}
f_{\omega}({\bf l},{\bf k}_b,{\bf q})\nonumber\\
+\frac{1}{({\bf l}-{\bf k}_a)^2}
\left( 
f_{\omega}({\bf l},{\bf k}_b,{\bf q}^2)
-\frac{{\bf k}_a^2 f_{\omega}({\bf k}_a,{\bf k}_b,{\bf q})}{{\bf l}^2+({\bf k}_a-{\bf l})^2}
\right)
\nonumber\\
+\frac{1}{({\bf l}-{\bf k}_a)^2}
\bigg( 
\frac{ ( {\bf k}_a-{\bf q}  )^2 {\bf l}^2 f_{\omega}({\bf l},{\bf k}_b,{\bf q}^2)}   
{ ({\bf l}-{\bf q})^2 {\bf k}_a^2 }\nonumber\\
-\frac{({\bf k}_a-{\bf q})^2 f_{\omega}({\bf k}_a,{\bf k}_b,{\bf q}^2)}
{ ({\bf l}-{\bf q})^2 ({\bf k}_a-{\bf l})^2 }
\bigg)
\bigg\}\,,
\label{eq:nbfkl}
\end{eqnarray} 
with $\bar{\alpha_s}=N_c \alpha_s/\pi$. This is the BFKL equation.
In the case of zero momentum transfer, ${\bf q^2}=0$,  Eq.~\ref{eq:nbfkl} becomes:
\begin{eqnarray}
\omega f_{\omega}({\bf k}_a,{\bf k}_b )=\delta^2({\bf k}_a-{\bf k}_b)\nonumber\\
+ \frac{\bar{ \alpha_s }}{2 \pi}   \int   
\frac{d^2{\bf l}}{({\bf l}-{\bf k}_a)^2}
\left( f_{\omega}({\bf l},{\bf k}_b )
-\frac{{\bf k}_a^2 f_{\omega}({\bf k}_a,{\bf k}_b )}{{\bf l}^2+({\bf k}_a-{\bf l})^2}
\right).
\label{eq:nbfklq0}
\end{eqnarray} 

The impossible task of summing an infinite number of integrals, 
each one with a $(i+1)$-body phase space if its previous has
an $i$-body phase space turns into finding a way to solve
Eq.~\ref{eq:nbfkl}.
We can rewrite the BFKL equation in a more symbolic form as
\begin{eqnarray}
\omega f_{\omega}({\bf k}_a,{\bf k}_b )=\delta^2({\bf k}_a-{\bf k}_b)
+ \int   d^2{\bf l}\,\,\mathcal{K}({\bf k}_a,{\bf l})\,f_{\omega}({\bf l},{\bf k}_b)\,,
\label{eq:nbfklsimpl}
\end{eqnarray} 
where $\mathcal{K}({\bf k}_a,{\bf l})$ is the BFKL kernel:
\begin{eqnarray}
\mathcal{K}({\bf k}_a,{\bf l})\,=\,
\underbrace{2 \epsilon(-{\bf k}^2)\,\delta^2({\bf k}_a-{\bf l})}_{\mathcal{K}_{virt}}
\,+{\underbrace{\frac{N_c \alpha_s}{\pi^2}\,\frac{1}{({\bf k}_a-{\bf k}_b)^2}}_{\mathcal{K}_{real}}}\,.
\label{eq:nkernel}
\end{eqnarray} 
$\mathcal{K}_{\mathrm{virt}}$ and $\mathcal{K}_{\mathrm{real}}$ are the parts of the kernel that correspond to
the virtual and real corrections respectively. 

Solving the BFKL equation will provide us with
the  BFKL gluon Green's function from which we can reconstruct the imaginary part 
of the amplitude 
for $q q$-scattering in two steps.
First, we need to perform the inverse Mellin transform to return to to $s$ space:
\begin{eqnarray}
f(s,{\bf k}_a,{\bf k}_b ,{\bf q})\,=\,
\frac{1}{2\pi i} \int_{c- i \infty}^{c +i \infty}
d\omega\,\left(\frac{s}{s_0}\right)^{\omega}\, f_{\omega}({\bf k}_a,{\bf k}_b,{\bf q} )
\label{eq:ninvmel}
\end{eqnarray} 
and subsequently we need to integrate over the ${\bf k}_a$
and ${\bf k}_b$ momenta of the reggeized gluons:
\begin{eqnarray}
\mathcal{A}_{\mathrm{singlet}}(s,t)=i (8 \pi \alpha_s)^2 \,s\,
\frac{N_c^2-1}{4 N_c^2}\,\int\,\frac{d^2{\bf k}_a}{(2
  \pi)^2}\,\frac{d^2{\bf k}_b}{(2 \pi)^2}\,
\frac{f(s,{\bf k}_a,{\bf k}_b,{\bf q} )}{{\bf k}_b^2({\bf k}_a-{\bf q})^2}\,,
\label{eq:nast}
\end{eqnarray} 
where we kept the color factor for $q q$-scattering explicit.
\begin{figure}
\centerline{\epsfig{file=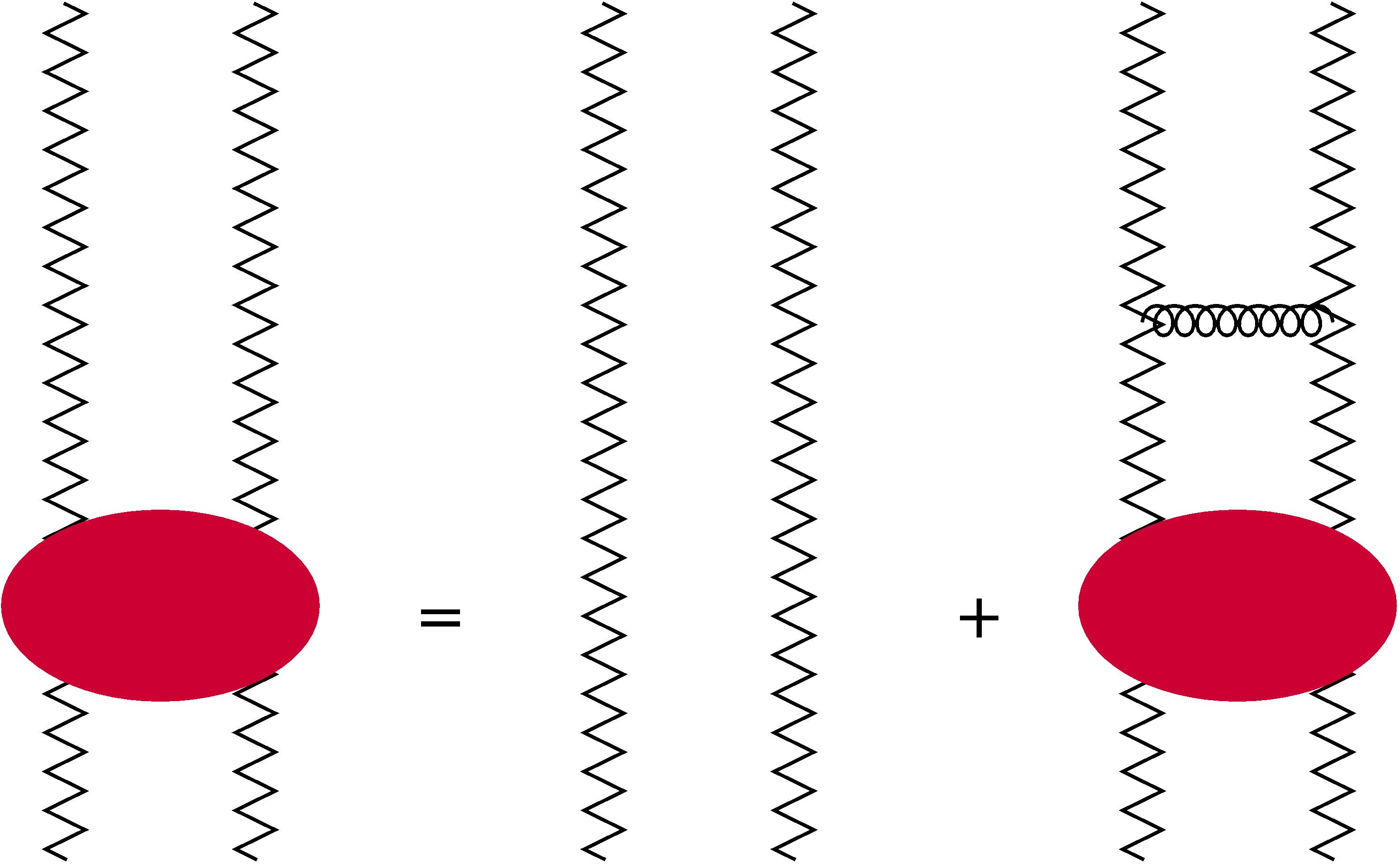,height=5cm}}
\caption{Graphical depiction of the BFKL equation.}
\label{fig:nbfklg}
\end{figure}

The amplitude in Eq.~\ref{eq:nast} is the amplitude for scattering via
a perturbative Pomeron exchange. If nature were to follow the BFKL
dynamics, or more precisely, in the kinematical limit where BFKL dynamics
is dominant and describes fully the perturbative QCD picture,
the interaction between two quarks would be the
outcome of summing all possible ladder diagrams with $n$-rungs, $n \rightarrow \infty$,
and this would be the equivalent of saying that the two quarks exchange a Pomeron.
This is obviously not a definition of the Pomeron but describes a good deal of how 
to perceive it in an intuitive manner.

The BFKL kernel in Eq.~\ref{eq:nkernel} is infrared finite, 
$\mathcal{K}_{\mathrm{real}}$ and $\mathcal{K}_{\mathrm{virt}}$
are both singular but their divergencies cancel one against the other. The amplitude though
is still infrared divergent due to the gluon propagators $\frac{1}{{\bf k}_b^2}$ 
and $\frac{1}{({\bf k}_a-{\bf q})^2}$. In practice, the quarks (or scattering gluons for that matter)
are not on mass-shell as we presupposed in the discussion of the BFKL equation so far.
In physical processes, as for example in
hadron hadron collisions at the LHC, the Pomeron couples to
off shell partons inside a hadron. To account for the hadronic structure, 
we need to introduce the notion of the impact factor,
$\Phi$, which is practically the coupling of the Pomeron to the hadron.
\begin{figure}
\centerline{\epsfig{file=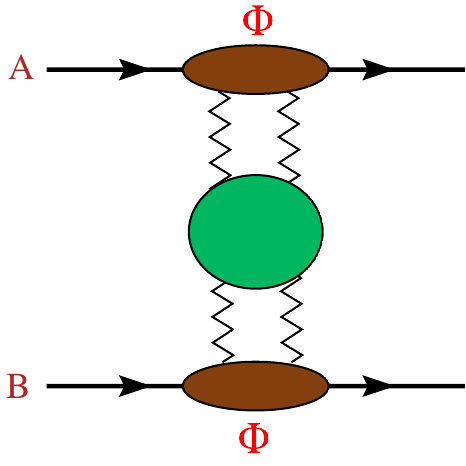,height=6cm}}
\caption{High energy hadron-hadron scattering.
The interaction factorizes into the process independent part which is the
BFKL gluon Green's function (green blob) and the its effective couplings to the
scattering projectiles, the impact factors (brown blobs).}
\label{fig:nhadronn}
\end{figure}
Then a hadronic
elastic amplitude between hadrons $A$ and $B$ (Fig.~\ref{fig:nhadronn}) will be written as
\begin{equation}
\mathcal{A}(s,t)\,=\,i\,s\,\mathcal{C}\,\int\,\frac{d^2{\bf k}_a}{(2 \pi)^2}\,
\frac{d^2{\bf k}_b}{(2 \pi)^2}\,
\Phi_A({\bf k}_a,{\bf q})\,
\frac{f(s,{\bf k}_a,{\bf k}_b,{\bf q})}{{\bf k}_b^2({\bf k}_a-{\bf q})^2}
\Phi_B({\bf k}_b,{\bf q})\,,
\label{eq:nfactH}
\end{equation} 
where $\mathcal{C}$ accounts for the colour factor of the process\footnote{For example, $\mathcal{C} =
(N_c^2-1)/4 N_c^2$ for $q q$-scattering}
and the quantities $\Phi_A$ and $\Phi_B$ are the hadron impact factors for 
the hadrons $A$ and $B$. Whenever we have a Pomeron exchange in
particle scattering, we need also to include in the analysis the
impact factors for each of the scattering parts.
In general, impact factors are process dependent object and mostly
of non perturbative nature and thus non-calculable and
subjects to modelling. Still, there has been quite significant effort
by the community to calculate perturbative impact factors to 
NLO~\cite{Fadin:1999df,Ciafaloni:2000sq,Fadin:1999de,Ivanov:2004pp,
Ivanov:2005xc,Bartels:2002yj,Bartels:2001ge,Caporale:2011cc,Ivanov:2012ms,
Ivanov:2012iv,Chachamis:2012cc,Chachamis:2012mw,
Bartels:2002uz,Bartels:2004bi,Bartels:2001mv,Bartels:2000gt,
Chachamis:2006zz,Balitsky:2012bs,Balitsky:2010ze,Chachamis:2013kra,Chachamis:2014sba,
Chachamis:2013bwa}. 
Nevertheless, all impact factors
have to share a very important universal behavior, i.e. 
they become zero in the limits
\begin{eqnarray}
\Phi({\bf k},  {\bf q})\Big|_{{ \bf k} \to  0}^{{\bf  k}-{\bf q} \to 0}\, \to 0 \,.
\end{eqnarray}  
and they regulate thus the infrared divergencies
of Eq.~\ref{eq:nfactH} which exactly
appear in these limits.

We can rewrite Eq.~\ref{eq:nbfklsimpl} as
\begin{equation}
\omega F\,=\,{\it 1}\!\!{\it I} + \mathcal{K}\;\otimes\;F,
\end{equation}
with $ \mathcal{K}$ being the BFKL kernel as in Eq.~\ref{eq:nkernel},
and  attempt to diagonalise the BFKL equation 
by finding the eigenfunctions $\phi_a$ of the kernel $\mathcal{K}$
\begin{equation}
\mathcal{K}\,\otimes\,\phi_a\, = \,\omega_a \phi_a\,.
\end{equation}
If $\theta$ is the azimuthal angle of the momenta,
then the eigenfunctions can be written as:
\begin{equation}
\phi_{n \nu}(|{\bm k}|,\theta)\, = \,\frac{1}{\pi \sqrt{2}}({\bm k}^2)^{-\frac{1}{2}+i \nu}\,e^{i n \theta}\,.
\end{equation}
The high energy
behavior of the total cross section is determined 
by the behavior of the angular averaged kernel 
(averaged over the
azimuthal angle between $\bm{k}_a$ and $\bm{k}_b$) and then 
$(\bm{k}^2)^{\gamma-1}$ can be used as eigenfunctions such that:
\begin{align}
\int d^{2}k \,\mathcal{K}(\bm{k}_a,\bm{k}) (\bm{k}^2)^{\gamma-1} =
\frac{N_c \as}{\pi} \,\chi_0(\gamma)(\bm{k}_a^2)^{\gamma-1}
\end{align} 
with eigenvalues
\begin{align*}
\omega_n(\gamma) = \frac{ \alpha_s N_c}{\pi}
\left( 2 \psi(1) - \psi(\gamma +\frac{n}{2}) - \psi(1-\gamma+\frac{n}{2}) \right),\:\:\:
\psi(\gamma)=\Gamma'(\gamma)/\Gamma(\gamma)
\end{align*}
and $\gamma = 1/2 + i \nu$.
The set of eigenfunctions is complete with
$\nu$ taking real values between $-\infty$ and $\infty$ . 
The solution can therefore be expressed using the expansion on the  eigenfunctions
and reads
\begin{equation}
f({\bf k}_a, {\bf k}_b, Y) = \frac{1}{\pi {\bf k}_a {\bf k}_b} \sum_{n=-\infty}^{\infty} \int
\frac{d\omega}{2 \pi i} e^{\omega Y} \int \frac{d\gamma}{2 \pi i} 
\left(\frac{{\bf k}_a^2}{{\bf k}_b^2}\right)^{\gamma-\frac{1}{2}}
\frac{e^{i n \theta}}{\omega - \omega_n(\alpha_s, \gamma)},
\label{eq:nrapsol}
\end{equation}
where $Y = \ln\left(\frac{s}{s_0}\right)$ is 
the rapidity interval between ${\bf k}_a$ and ${\bf k}_b$.
Eq.~\ref{eq:nrapsol} makes apparent the distinct  power-like 
growth with energy prediction
within the BFKL dynamics that characterises the  behavior of the cross
sections  at large energies. The relevant term here is $e^{\omega Y}$.

So far, we have encountered a number of important
features of the BFKL resummation program  all seen
at leading logarithmic
accuracy. At next-to-leading logarithmic approximation (NLLA),
it turns out that the reggeization of the gluon still holds which is a key point.
It means that one can use the leading order form of BFKL equation
changing only the kernels and
the eigenvalues~\cite{Fadin:1998py}. We will not discuss
in any detail the NLO BFKL equation here. We will only sketch 
the origin of the terms $\as(\as \ln s)^n$ and we will mention a couple
of important points for BFKL phenomenology.

The NLO\footnote{As mentioned in the introduction,
in the field, there is an interchangeability 
between the terms `NLL' and `NLO'. We will follow
the practise here with the assurance 
that by now the context makes clear what one really means.} 
corrections stem from two
different kinematical configurations.
In MRK, the next-to-leading order corrections for the gluon Regge
trajectory as well as the virtual corrections
to the Reggeon-Reggeon-$g$ vertex have to be included.
The reggeized gluon trajectory has to be calculated at two-loop
approximation, $\epsilon^{(2)}$ \cite{nomega2}, whereas, the real part of the kernel,
$\mathcal{K}_{\mathrm{real}}$ gets contributions from one-loop level gluon production  \cite{nRRG2}. 

\begin{figure}
\centerline{\epsfig{file=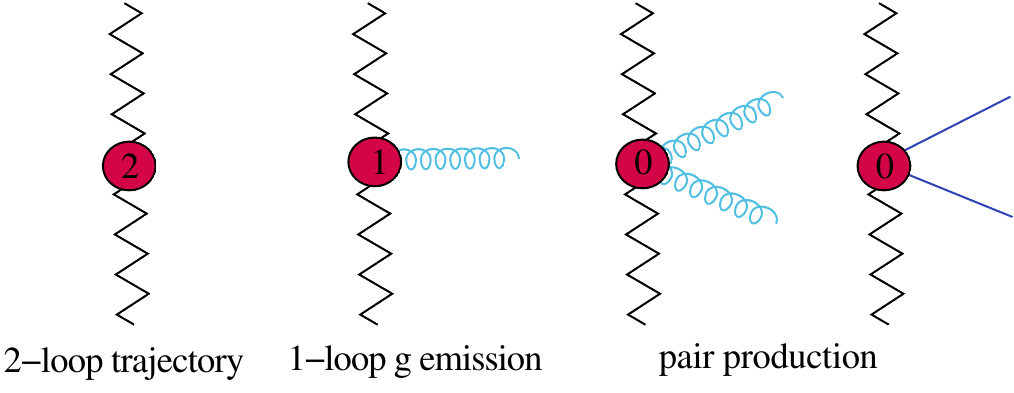,height=4.5cm}}
\caption{Schematic representation of the configurations
that contribute to the NLLA approximation.}
\label{fig:nNLLA}
\end{figure}

One can also obtain a term of the type $\as(\as \ln s)^n$ 
starting from an amplitude at LLA and after losing a relative $\ln s$ term. 
We saw that the key feature that generates
these logarithmic terms 
is the strong ordering of the emitted gluons in rapidity space.
Thus, if we allow for a state in which
two  of the emitted particles are close in rapidity, 
we are in the 
Quasi-Multi-Regge-kinematics (QMRK) where
Eq.~\ref{eq:nMRK} still holds with the exception of a pair of particles.
The pair can be a pair of gluons or a $q\bar q$
pair  \cite{nRRGG,nRRQQ} (Fig.~\ref{fig:nNLLA}).

The calculation of the NLLA corrections was an difficult task
that took almost a decade~\cite{Fadin:1998py,Ciafaloni:1998gs}.
After its completion, it turned out that the NLLA corrections compared to
the LLA term were very large
questioning the convergence itself of the perturbative expansion
in terms of $(\alpha_s \ln s)$. The
problem has its origin to the fact that since the
transverse momenta of the emitted gluons are not restricted, 
there can be final states
in which the transverse momenta 
are strongly ordered. This in turn means that large logarithms of transverse
momenta (collinear logarithms) can be present and render
the expansion in $(\as \ln s)$ terms unstable. Therefore,
one needs to perform a complete collinear  resummation of these
large logarithms in order to stabilize the convergence of the expansion~\cite{nResum1}.

\section{BFKL phenomenology at the LHC}

In the  past
thirty years, a number of probes of BFKL physics have been
proposed for different collider environments. 
Actually, BFKL phenomenology had its first major flourish in the nineties,
especially after HERA at DESY started producing data for the proton
structure function $F_2$ that were showing a power-like rise with decreasing
$x$, the Bjorken scaling variable. Since the early HERA days,
much of the progress seen 
on more formal theoretical issues regarding the BFKL formalism
was driven from a need to compare against experimental measurements.
Nevertheless, the absence of a clear
signal that would only be described by BFKL physics and by nothing else
was a drawback. Despite the big progress in the field,
most of the studies
we still have are beyond LO but only a few
calculations provide full NLO accuracy estimates within the BFKL framework.

\begin{figure}
\centerline{\epsfig{file=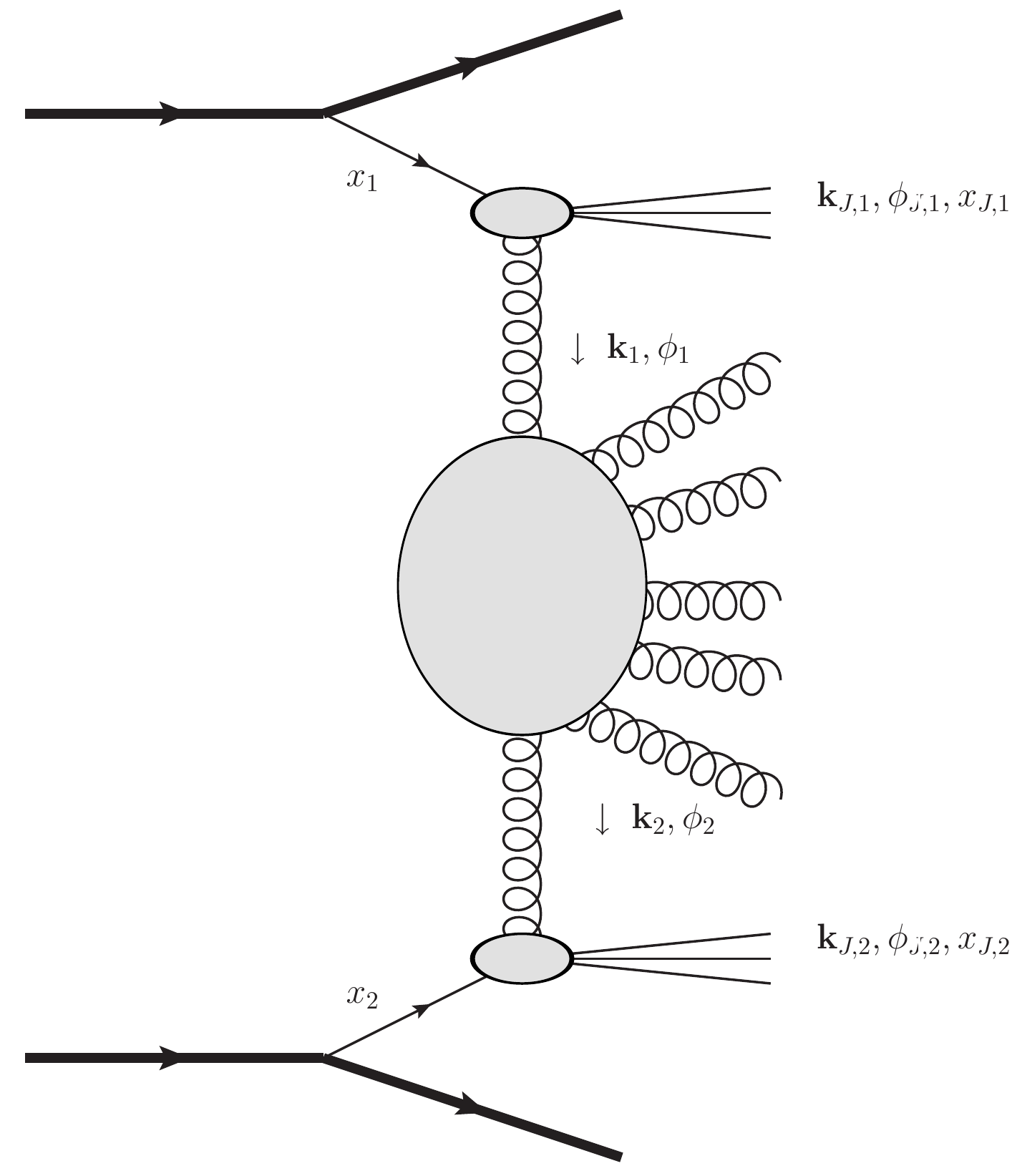, height=10cm}}
\caption{The kinematics of Mueller-Navelet jets. Figure taken from Ref.~\cite{Ducloue:2013wmi}.}
\label{fig:nMN}
\end{figure}

Nowadays, the general consensus
is that one should apply the BFKL formalism to processes that have two
hard scales at the two ends of the BFKL ladder that are of the same magnitude.
Otherwise, if there is  strong
ordering in the transverse momentum of the two scales, DGLAP~\cite{Dokshitzer:1977sg,
Gribov:1972ri, Altarelli:1977zs} logarithms appear and
BFKL is not any more the only relevant framework. 
A very strict list of probes would include the processes 
$\gamma^* \gamma^* \rightarrow$ hadrons in a $e^+ e^-$ collider,
forwards jets in Deep Inelastic Scattering (DIS) at HERA,
Mueller-Navelet jets and Mueller-Tang jets at hadron colliders (Tevatron, LHC).
We do not include in the list the $F_2$ behavior in DIS which is also
driven by non-perturbative physics.
From the list of probes above,  Mueller-Navelet jets \cite{Mueller:1986ey}
 is the observable that
has received most of the theoretical attention in recent years as
the process can be studied experimentally at the LHC. In the following,
we will focus on recent Mueller-Navelet studies and we
will also review the comparison with experimental data.

The initial idea behind considering the Mueller-Navelet jets cross section as
a probe for BFKL physics is the following: in a hadron collider, 
let us assume that two partons interact (one from each hadron)
such that in the final state we find a forward and backward jet
of similar and sizeable $p_T$. Then these can be the hard scales attached
to the two ends of a BFKL ladder and any collinear (DGLAP) logarithms
are suppressed in the evolution from one jet to the other. The main
contribution to this process on the partonic level
then would come from the BFKL logarithms
given that the two jets are well separated in rapidity.
The process is depicted in Fig.~\ref{fig:nMN}.

\begin{figure}
\centerline{\epsfig{file=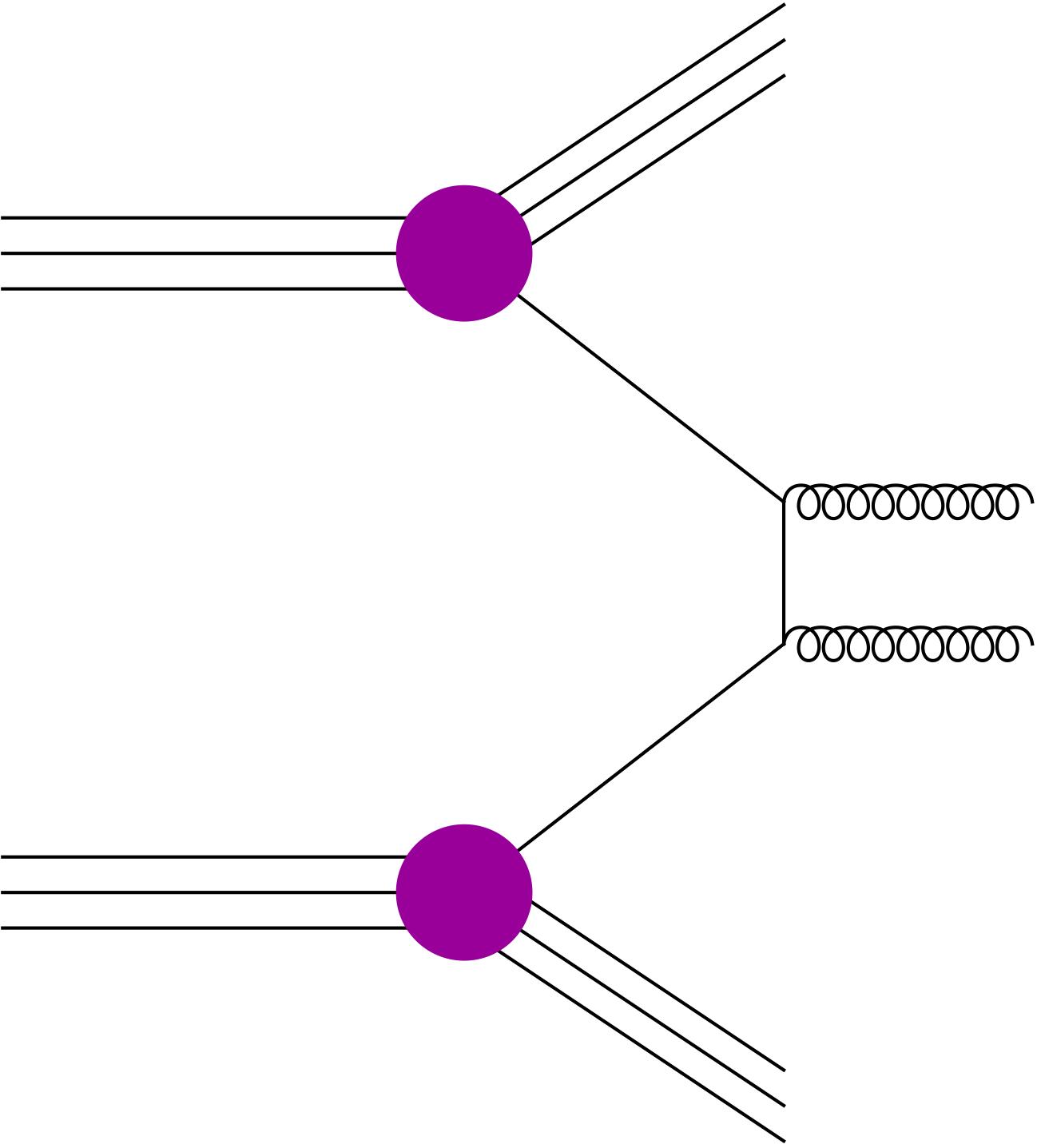, height=6cm}}
\caption{Diagrammatic tree level approximation for Mueller-Navelet jets.}
\label{fig:nMNtree}
\end{figure}

One would think that already at Tevatron, the aim would
be to see in the data
the power-like growth  with energy of the cross section characteristic for BFKL
dynamics. The problem with that though is that
 this growth is drown due to the rapidly falling PDFs in 
forward-backward dijet production with large rapidity
separation. For that reason, the main observable 
to be studied is the decorrelation in azimuthal angle 
 between the two tagged jets as a function of the rapidity separation~\cite{DelDuca:1993mn,Stirling:1994he,Orr:1997im,Kwiecinski:2001nh}.
 
\begin{figure}
\centerline{\epsfig{file=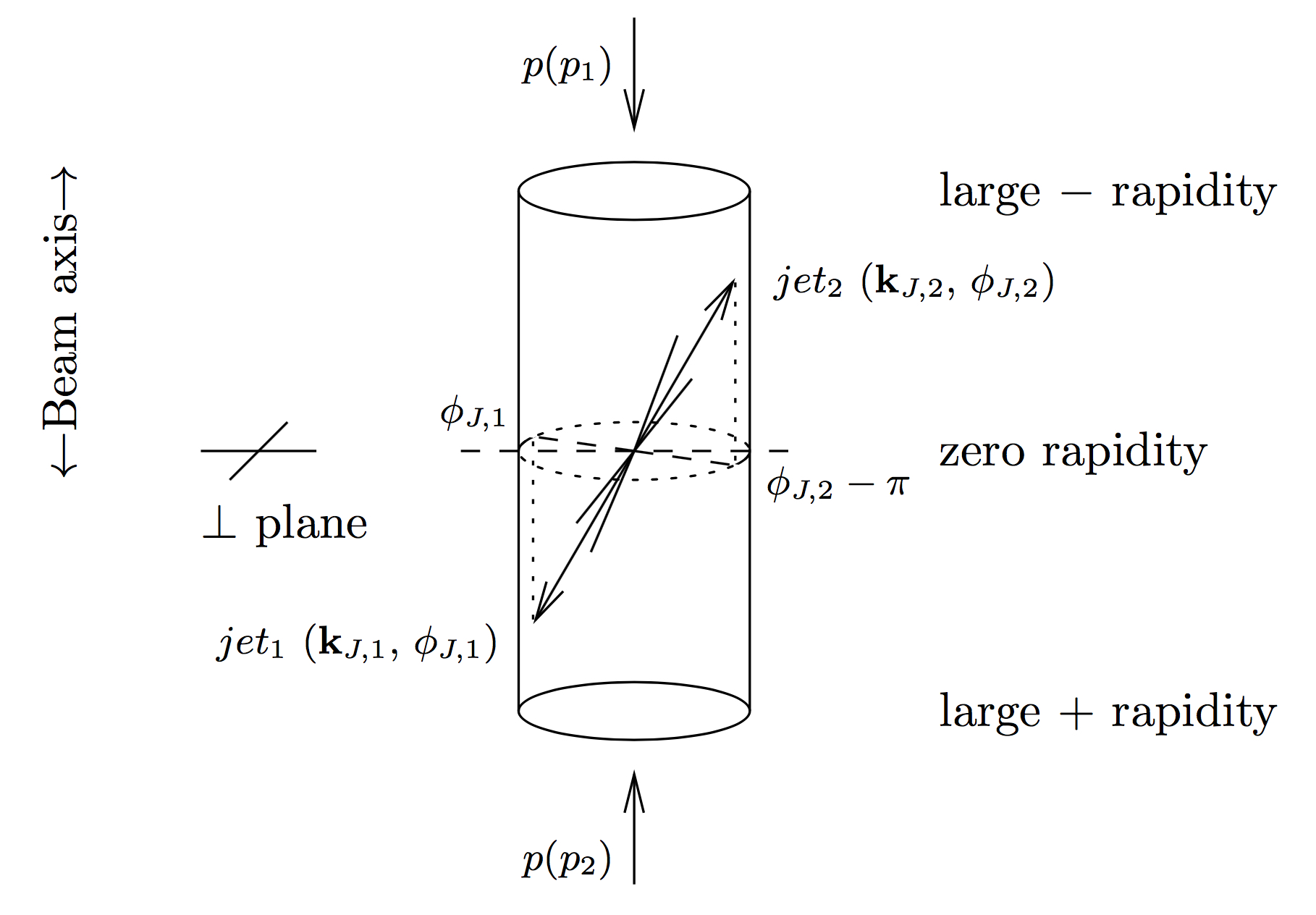, height=8cm}}
\caption{Tree level approximation for Mueller-Navelet jets in a collision setup.
Figure taken from Ref.~\cite{Colferai:2010wu}.}
\label{fig:nMN2010}
\end{figure}
\begin{figure}
\centerline{\epsfig{file=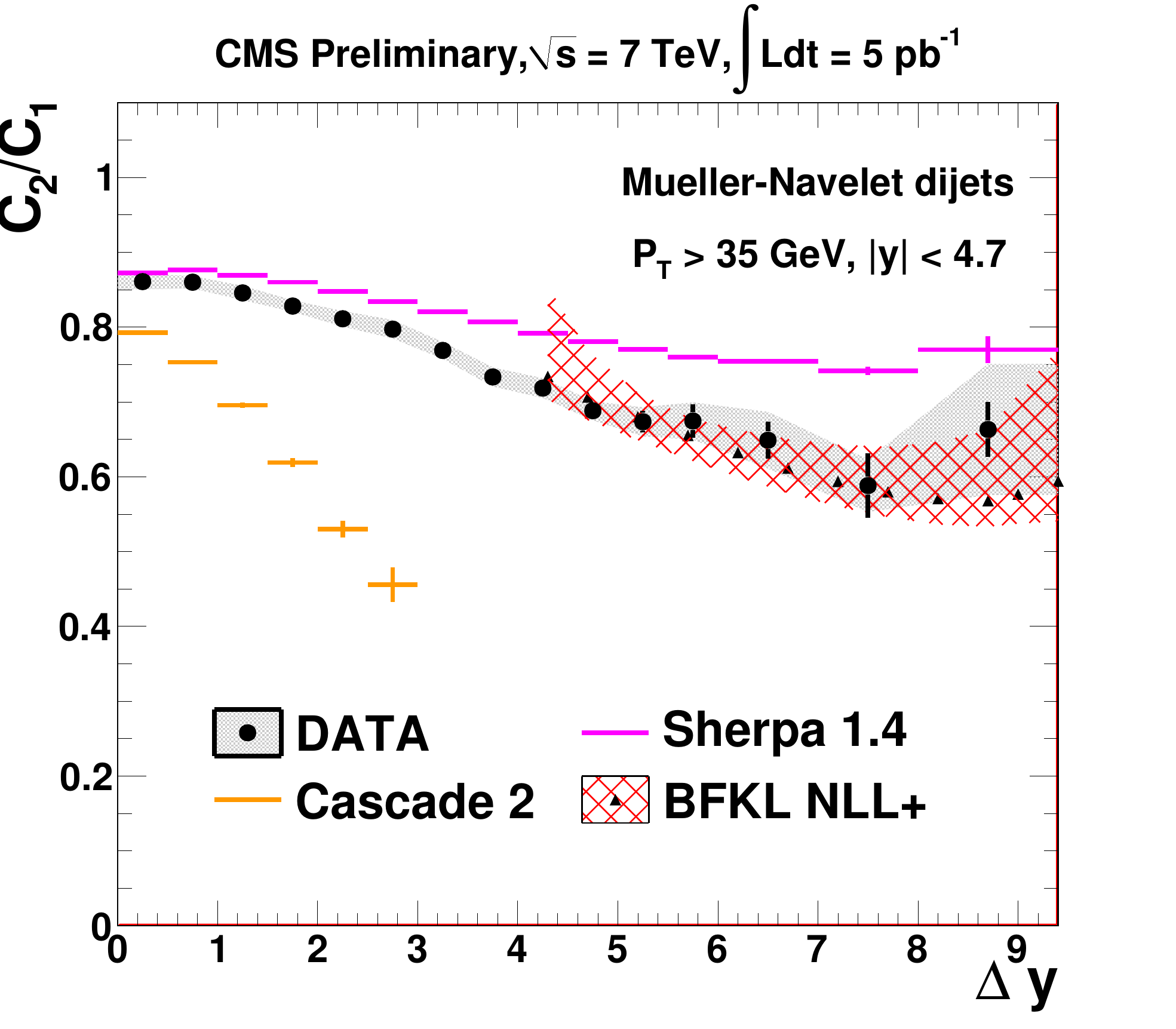, height=8cm}}
\caption{Ratio $C_2/C_1$ as a function of $\Delta y$ compared to various theory predictions.
Figure taken from Ref.~\cite{Safronov:2015bva}.}
\label{fig:nrat211}
\end{figure}

\begin{figure}
\centerline{\epsfig{file=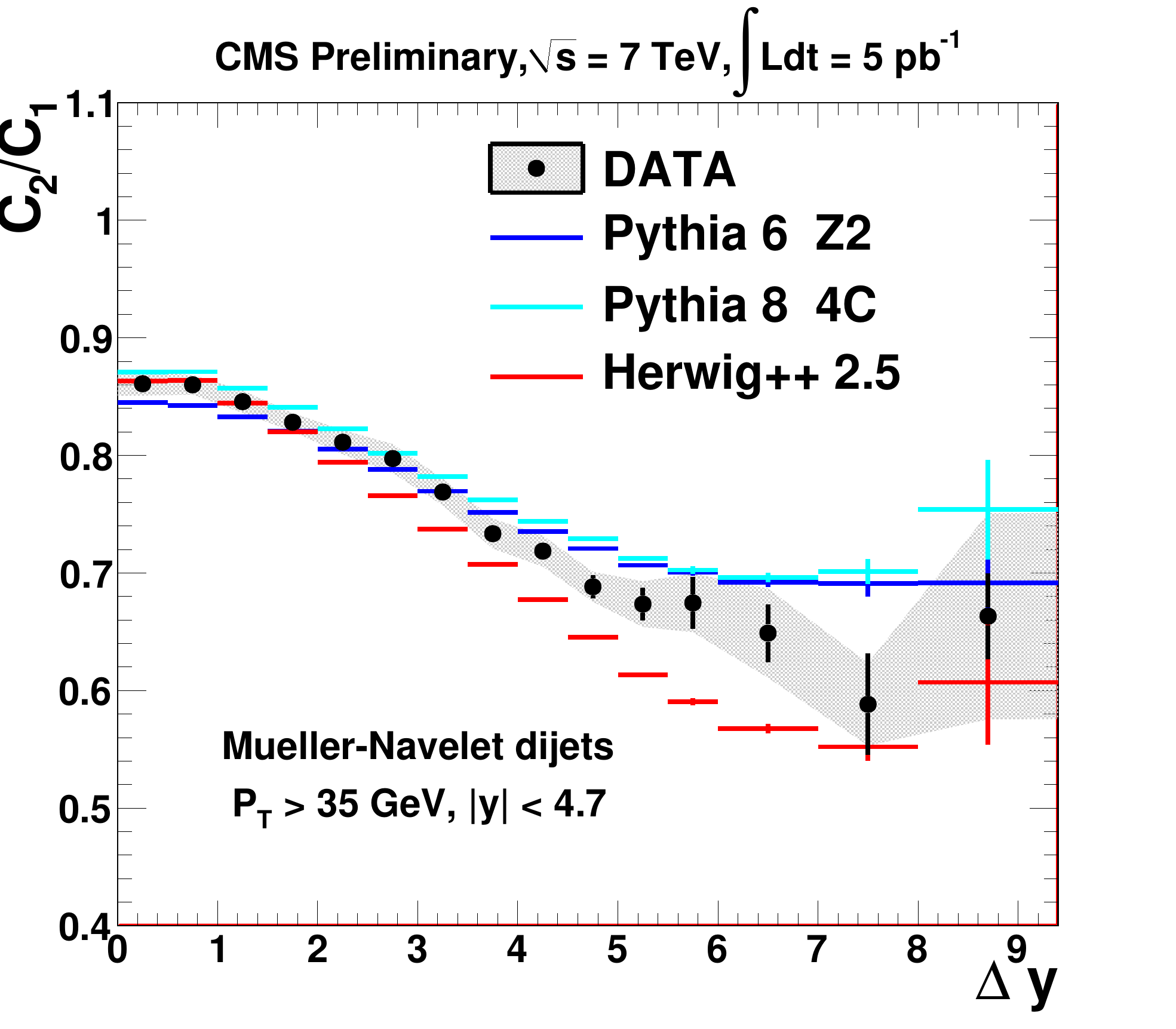, height=8cm}}
\caption{Ratio $C_2/C_1$ as a function of $\Delta y$ compared to various theory predictions.
Figure taken from Ref.~\cite{Safronov:2015bva}.}
\label{fig:nrat212}
\end{figure}
 
At tree level (Fig.~\ref{fig:nMNtree}), 
the produced jets have to be back-to-back due to energy-momentum
conservation: the partonic cross section is a $2 \rightarrow 2$ process.
As the partonic centre-of-mass energy increases though, the tree level
approximation is not a good approximation at all. One is bound to consider
extra real radiation in the final state which breaks the back-to-back
configuration of the two outmost jets. The larger the available energy,
the larger is the phase space and
more emissions need to be considered in order to describe more
accurately what really happens in the collider
and the more azimuthally decorrelated is the system of the forward-backward
jets. To measure the correlation, one projects the momenta of the two jets
on the transverse plane and calculates the average $cos(\Delta \phi)$, where
$\Delta \phi$ is defined as the difference of the angles of the two jets minus $\pi$,
$\Delta \phi = \phi_{J,1} - \phi_{J,2} - \pi$.
One can go further along these lines and compute the following moments:
$C_n = \langle cos(n \Delta \phi) \rangle$, where $n$ can be 1,2 or 3. 
In an effort
to minimise further any contamination from collinear logarithm, the ratios 
\begin{equation}
\frac{C_n}{C_m}= \frac{\langle cos(n \Delta \phi) \rangle}{\langle cos(m \Delta \phi) \rangle}
\end{equation}
have been proposed as better observables to probe BFKL dynamics
\cite{Vera:2006un,Vera:2007kn,Angioni:2011wj}.

At the moment, we have two groups with 
full NLO BFKL  predictions for Mueller-Navelet jet observables at LHC energies.
Both groups are using an
analytic approach  (as opposed to Monte Carlo studies)~\cite{
Ducloue:2013wmi,Colferai:2010wu,Ducloue:2013bva,Caporale:2014gpa,Celiberto:2015yba}. 
In their studies, they
compare and find good agreement with the average cosine ratios.
This agreement was summarized 
in the results of CMS on multijet correlations~\cite{Safronov:2015bva}
where Figs.~\ref{fig:nrat211} and~\ref{fig:nrat212} are taken from.
The success of BFKL physics to describe the data for the average
cosine ratios and the not so good performance of the standard
collinear tools is a very promising starting point while waiting
for relevant results from the second run of the LHC.

\begin{figure}
\centerline{\epsfig{file=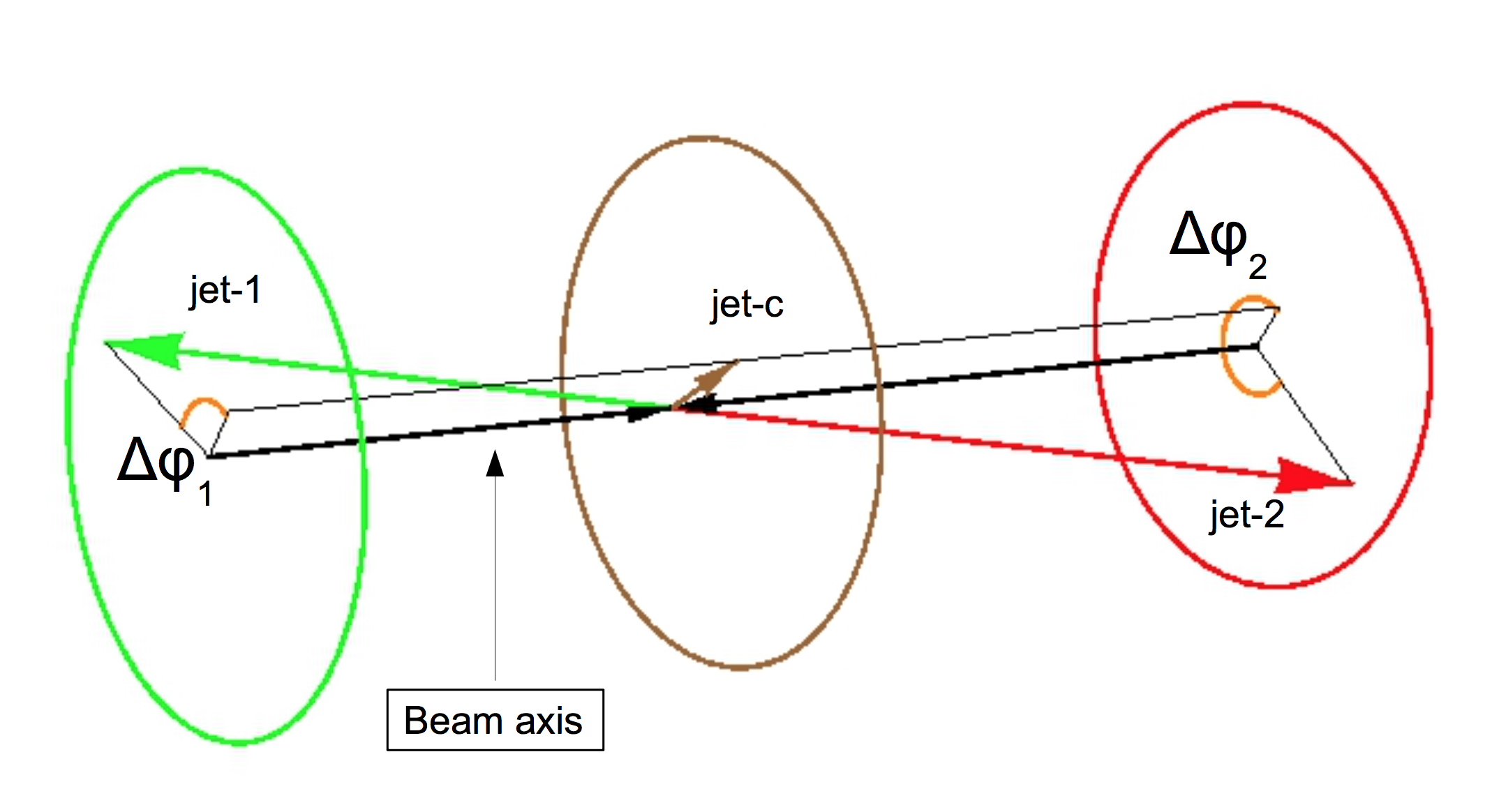, height=7cm}}
\caption{Kinematics of a 3-jet event.}
\label{fig:n3-jets}
\end{figure}

\begin{figure}
\centerline{\epsfig{file=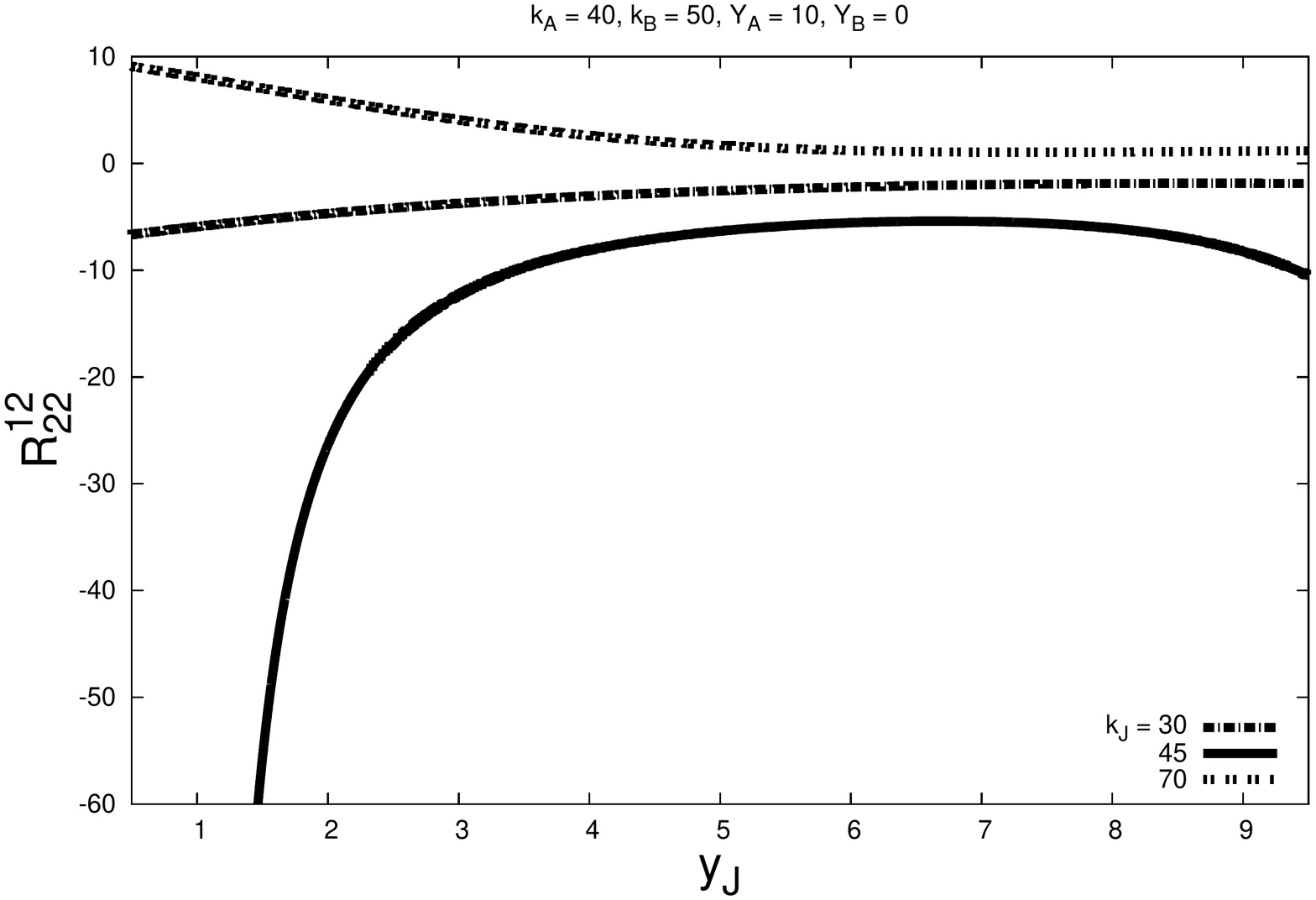, height=8cm}}
\caption{A study of the ratio $R^{12}_{22}$ as defined in Eq.~\ref{eq:nrr} for
fixed values of the $p_T$ of the two forward jets and three 
values of the $p_T$ of the tagged central jet, as a function of the rapidity of the central jet $y_J$.}
\label{fig:nr1222}
\end{figure}

Recently, new observables sensitive to BFKL dynamics
were proposed in the context of multijet production at the LHC~\cite{Caporale:2015vya,Caporale:2015int}.
The idea is to study events with two tagged forward-backward jets, 
separated by a large rapidity span, and also tag on a third jet\footnote{
In Ref.~\cite{Caporale:2015int}, two --more central in rapidity-- jets are tagged
but we restrict the discussion here to the 3-jet observables.} 
produced in the central region of rapidity, allowing for 
inclusive radiation in the remaining areas of the detectors.
A kinematical configuration can be seen in Fig~\ref{fig:n3-jets}.
The proposed distributions have a very 
different behavior to the ones characteristic 
of the Mueller-Navelet case. These new distributions 
are defined using the projections on the two relative 
azimuthal angles formed by each of the forward jets with the central jet,
$\Delta \phi_1 = \phi_1 - \phi_c -\pi$ and $\Delta \phi_1 = \phi_c - \phi_2 -\pi$.
The experimentally relevant observable is the mean 
value in the selected events of the two cosines of the azimuthal angle
differences, i.e. $\langle cos(M \Delta \phi_1)  cos(N \Delta \phi_2) \rangle$.
To eliminate again any collinear logarithm contamination, one can
form ratios and finally the observables are defined as:
\begin{equation}
\mathcal{R}^{M, N}_{P, Q} = \frac{\langle cos(M \Delta \phi_1)  cos(N \Delta \phi_2) \rangle}
{\langle cos(P \Delta \phi_1)  cos(Q \Delta \phi_2) \rangle},
\label{eq:nrr}
\end{equation}
where $M$, $N$, $P$ and $Q$ can be equal to 1 or 2.

In Fig.~\ref{fig:nr1222} one sees plotted the ratio $R^{12}_{22}$ after setting the
momentum of the forward jet to $k_A = 40$ GeV, the momentum of the backward jet
to $k_B = 50$ GeV and their rapidities to $Y_A = 10$ and $Y_B = 0$ respectively. 
For the transverse momentum of the central jet 
three values $k_J = 30, 45, 70$ GeV were chosen and the
rapidity of the central jet $y_J$ varies 
between the two rapidities of the forward-backward jets. The claim is that
these ratio distributions as defined in Eq.~\ref{eq:nrr}
are probing the fine structure of the QCD 
radiation in the high energy limit and one should
expect the LHC data to agree with the theoretical BFKL estimates
especially in the regions where $y_J$ is closer to $(Y_A-Y_B)/2$.
Apart from the analytic approach followed in Refs.~\cite{Caporale:2015vya,Caporale:2015int},
it would be very important to 
compare against the BFKL Monte Carlo code 
{\tt{BFKLex}}~\cite{Chachamis:2013rca,Caporale:2013bva,Chachamis:2012qw,Chachamis:2012fk,Chachamis:2011nz,Chachamis:2011rw,Chachamis:2015zzp,Chachamis:2015ico}
for these proposed ratio observables.

\section{Discussion}

The LHC has opened up a new era in particle physics. So far,
there is no clear signal for new physics and the SM seems to
secure even more its position as the best theory we have
to describe the fundamental interactions
(Gravity excluded). Despite that though, there is an awful lot we do not
know about the SM. If we exclude lattice works,
the only way we have at our disposal
to do calculations  is perturbation
theory. And clearly, knowing the first two-three terms of an expansion
to a function
does not give  a full insight to the function itself and its 
special properties. It only allows  to learn about the behavior of
the function in the small region where the expansion makes sense.
It remains to be seen whether the LHC era
will be an exciting time of new physics but even if not,
it should be the era in which 
we  learn and understand more about
the SM, especially more so if it surfaces at the end of
the day as the only fundamental theory available to describe 
consistently  experimental data. 

To that end, the role of phenomenology is crucial. We do not
calculate theoretical estimates and then compare to data sets
in order to fill out a checklist of processes. We do confront
our theory using the experiment because we want to understand
better our theory. We want to see whether different approaches
within the same fundamental model can reveal properties that were
previously masked. Phenomenology in modern
particle physics, apart from 
carrying  the responsibility of validating or falsifying a theory,
it should also shed light to corners of a valid theory 
that are not in plain view.

BFKL physics is connected to some very important and still
open issues within QCD and beyond.
Factorization theorems, the transition from the 
perturbative to the non-perturbative regime, the correct degrees of freedom
in high energies, the connection of QCD to the old Regge theory are few
examples. BFKL phenomenology should try to give answers to all these
important questions. Before that though, it needs to answer
the most pressing question: which is the rough collision energy threshold 
after which BFKL dynamics becomes --for the relevant kinematical configurations--,
if not dominant, at least the main player. Does the LHC reach beyond that threshold?
In that respect, to find the window of applicability for this formalism, more work 
is needed in identifying observables where the BFKL approach 
is distinct. We should define more exclusive
experimental quantities such that BFKL 
fits the measured data and all 
other possible approaches fail if we are already beyond the threshold at the LHC.
It remains to be seen whether
the second run of the LHC will be the time
of great progress for BFKL phenomenology.

\begin{flushleft}
{\bf \large Acknowledgements}
\end{flushleft}
The author would like to thank the organisers and in particular C. Royon
for the invitation, all the help and the warm hospitality.


\begin{thebibliography}{99}


\bibitem{Lipatov:1976zz} 
  L.~N.~Lipatov,
  Sov.\ J.\ Nucl.\ Phys.\  {\bf 23}, 338 (1976)
  [Yad.\ Fiz.\  {\bf 23}, 642 (1976)].
  
\bibitem{Fadin:1975cb} 
  V.~S.~Fadin, E.~A.~Kuraev and L.~N.~Lipatov,
  Phys.\ Lett.\ B {\bf 60}, 50 (1975).
  
\bibitem{Kuraev:1976ge} 
  E.~A.~Kuraev, L.~N.~Lipatov and V.~S.~Fadin,
  Sov.\ Phys.\ JETP {\bf 44}, 443 (1976)
  [Zh.\ Eksp.\ Teor.\ Fiz.\  {\bf 71}, 840 (1976)].
  
\bibitem{Balitsky:1978ic} 
  I.~I.~Balitsky and L.~N.~Lipatov,
  Sov.\ J.\ Nucl.\ Phys.\  {\bf 28}, 822 (1978)
  [Yad.\ Fiz.\  {\bf 28}, 1597 (1978)].
  
  
\bibitem{Fadin:1998py} 
  V.~S.~Fadin and L.~N.~Lipatov,
  Phys.\ Lett.\ B {\bf 429}, 127 (1998)
  [hep-ph/9802290].
  
\bibitem{Ciafaloni:1998gs} 
  M.~Ciafaloni and G.~Camici,
  Phys.\ Lett.\ B {\bf 430}, 349 (1998)
  [hep-ph/9803389].



\bibitem{nbarone}
V. Barone and E predazzi,
{\it High-Energy Particle Difraction}, Springer 2002.


\bibitem{nforshawross}
J. R. Forshaw and D. A. Ross,
{\it Quantum Chromodynamics and the Pomeron}, Cambridge 1997
(Cambridge lecture notes in physics 9).


\bibitem{nqcdlev}
B. L. Ioffe, V. S. Fadin and L. N. Lipatov,
{\it Quantum Chromodynamics. Perturbative and Nonperturbative Aspects}, Cambridge 2014
(Cambridge Monographs on Particle Physics, Nuclear Physics and Cosmology 30).

\bibitem{DelDuca:1995hf} 
  V.~Del Duca,
  hep-ph/9503226.

\bibitem{Salam:1999cn} 
  G.~P.~Salam,
  Acta Phys.\ Polon.\ B {\bf 30}, 3679 (1999)
  [hep-ph/9910492].

\bibitem{Fadin:1998sh} 
  V.~S.~Fadin,
  hep-ph/9807528.





\bibitem{Fadin:1999df} 
  V.~S.~Fadin, R.~Fiore, M.~I.~Kotsky and A.~Papa,
  Phys.\ Rev.\ D {\bf 61}, 094006 (2000)
  [hep-ph/9908265].

\bibitem{Ciafaloni:2000sq} 
  M.~Ciafaloni and G.~Rodrigo,
  JHEP {\bf 0005}, 042 (2000)
  [hep-ph/0004033].


\bibitem{Fadin:1999de} 
  V.~S.~Fadin, R.~Fiore, M.~I.~Kotsky and A.~Papa,
  Phys.\ Rev.\ D {\bf 61}, 094005 (2000)
  [hep-ph/9908264].

\bibitem{Ivanov:2004pp} 
  D.~Y.~Ivanov, M.~I.~Kotsky and A.~Papa,
  Eur.\ Phys.\ J.\ C {\bf 38}, 195 (2004)
  [hep-ph/0405297].

\bibitem{Ivanov:2005xc} 
  D.~Y.~Ivanov, M.~I.~Kotsky and A.~Papa,
  Nucl.\ Phys.\ Proc.\ Suppl.\  {\bf 146}, 117 (2005).

\bibitem{Bartels:2002yj} 
  J.~Bartels, D.~Colferai and G.~P.~Vacca,
  Eur.\ Phys.\ J.\ C {\bf 29}, 235 (2003)
  [hep-ph/0206290].

\bibitem{Bartels:2001ge} 
  J.~Bartels, D.~Colferai and G.~P.~Vacca,
  Eur.\ Phys.\ J.\ C {\bf 24}, 83 (2002)
  [hep-ph/0112283].

\bibitem{Caporale:2011cc} 
  F.~Caporale, D.~Y.~Ivanov, B.~Murdaca, A.~Papa and A.~Perri,
  JHEP {\bf 1202}, 101 (2012)
  [arXiv:1112.3752 [hep-ph]].
  
\bibitem{Ivanov:2012ms} 
  D.~Y.~Ivanov and A.~Papa,
  JHEP {\bf 1205}, 086 (2012)
  [arXiv:1202.1082 [hep-ph]].

\bibitem{Ivanov:2012iv} 
  D.~Y.~Ivanov and A.~Papa,
  JHEP {\bf 1207}, 045 (2012)
  [arXiv:1205.6068 [hep-ph]].


\bibitem{Chachamis:2012cc} 
  G.~Chachamis, M.~Hentschinski, J.~D.~Madrigal Martinez and A.~S.~Vera,
  Phys.\ Rev.\ D {\bf 87}, no. 7, 076009 (2013)
  [arXiv:1212.4992].
  
\bibitem{Chachamis:2012mw} 
  G.~Chachamis, M.~Hentschinski, J.~D.~Madrigal Martinez and A.~Sabio Vera,
  Phys.\ Part.\ Nucl.\  {\bf 45}, no. 4, 788 (2014)
  [arXiv:1211.2050 [hep-ph]].

\bibitem{Bartels:2002uz} 
  J.~Bartels, D.~Colferai, S.~Gieseke and A.~Kyrieleis,
  Phys.\ Rev.\ D {\bf 66}, 094017 (2002)
  [hep-ph/0208130].

\bibitem{Bartels:2004bi} 
  J.~Bartels and A.~Kyrieleis,
  Phys.\ Rev.\ D {\bf 70}, 114003 (2004)
  [hep-ph/0407051].
  
\bibitem{Bartels:2000gt} 
  J.~Bartels, S.~Gieseke and C.~F.~Qiao,
  Phys.\ Rev.\ D {\bf 63}, 056014 (2001)
  [Phys.\ Rev.\ D {\bf 65}, 079902 (2002)]
  [hep-ph/0009102].


\bibitem{Bartels:2001mv} 
  J.~Bartels, S.~Gieseke and A.~Kyrieleis,
  Phys.\ Rev.\ D {\bf 65}, 014006 (2002)
  [hep-ph/0107152].


\bibitem{Chachamis:2006zz} 
  G.~Chachamis and J.~Bartels,
  PoS DIFF {\bf 2006}, 026 (2006).

\bibitem{Balitsky:2012bs} 
  I.~Balitsky and G.~A.~Chirilli,
  Phys.\ Rev.\ D {\bf 87}, no. 1, 014013 (2013)
  [arXiv:1207.3844 [hep-ph]].

\bibitem{Balitsky:2010ze} 
  I.~Balitsky and G.~A.~Chirilli,
  Phys.\ Rev.\ D {\bf 83}, 031502 (2011)
  [arXiv:1009.4729 [hep-ph]].


\bibitem{Chachamis:2013kra} 
  G.~Chachamis, M.~Deak and G.~Rodrigo,
  PoS EPS {\bf -HEP2013}, 092 (2013)
  [arXiv:1310.7763 [hep-ph]].

\bibitem{Chachamis:2014sba} 
  G.~Chachamis, M.~Deak and G.~Rodrigo,
  PoS DIS {\bf 2014}, 085 (2014).

\bibitem{Chachamis:2013bwa} 
  G.~Chachamis, M.~Deak and G.~Rodrigo,
  JHEP {\bf 1312}, 066 (2013)
  [arXiv:1310.6611 [hep-ph]].



\bibitem{nomega2}
V.~S.~Fadin, 
Zh. Eksp. Teor. Fiz. Pis'ma {\bf 61} (1995) 342;
V.~S.~Fadin, R.~Fiore and A.~Quartarolo,
Phys.\ Rev. {\bf D53} (1996) 2729;
V.~S.~Fadin, M.~I.~Kotsky, Yad. Fiz. {\bf 59}(6) (1996) 1;
V.~S.~Fadin, M.~I.~Kotsky and R.~Fiore,
Phys.\ Lett. {\bf B359} (1995) 181.


\bibitem{nRRG2}
V.~S.~Fadin and L.~N.~Lipatov,
Nucl.\ Phys. {\bf B406} (1993) 259;
V.~S.~Fadin, R.~Fiore and A.~Quartarolo,
Phys.\ Rev. {\bf D50} (1994) 5893

\bibitem{nRRGG}
V.~S.~Fadin, M.~I.~Kotsky and L.~N.~Lipatov,
Phys.\ Lett. {\bf B415} (1997) 97.

\bibitem{nRRQQ}
S.~Catani, M.~Ciafaloni and F.~Hautmann,
Phys.\ Lett.  {\bf B242} (1990) 97;
Nucl.\ Phys.  {\bf B366} (1991) 135;
G.~Camici and M.~Ciafaloni,
Phys.\ Lett.  {\bf B386} (1996) 341;
V.~S.~Fadin, R.~Fiore, A.~Flachi and M.~I.~Kotsky,
Phys.\ Lett.  {\bf B422} (1998) 287

\bibitem{nResum1}
M.~Ciafaloni, D.~Colferai and G.~P.~Salam,
Phys.\ Rev. {\bf D60} (1999) 114036


\bibitem{Dokshitzer:1977sg} 
  Y.~L.~Dokshitzer,
  Sov.\ Phys.\ JETP {\bf 46}, 641 (1977)
  [Zh.\ Eksp.\ Teor.\ Fiz.\  {\bf 73}, 1216 (1977)].
  
\bibitem{Gribov:1972ri} 
  V.~N.~Gribov and L.~N.~Lipatov,
  Sov.\ J.\ Nucl.\ Phys.\  {\bf 15}, 438 (1972)
  [Yad.\ Fiz.\  {\bf 15}, 781 (1972)].
  
\bibitem{Altarelli:1977zs} 
  G.~Altarelli and G.~Parisi,
  Nucl.\ Phys.\ B {\bf 126}, 298 (1977).


\bibitem{Mueller:1986ey}
  A.~H.~Mueller and H.~Navelet,
  Nucl.\ Phys.\ B {\bf 282} (1987) 727.


\bibitem{DelDuca:1993mn}
  V.~Del Duca and C.~R.~Schmidt,
  Phys.\ Rev.\ D {\bf 49} (1994) 4510
  [hep-ph/9311290].
  
\bibitem{Stirling:1994he}
  W.~J.~Stirling,
  Nucl.\ Phys.\ B {\bf 423} (1994) 56
  [hep-ph/9401266].
  
\bibitem{Orr:1997im}
  L.~H.~Orr and W.~J.~Stirling,
  Phys.\ Rev.\ D {\bf 56} (1997) 5875
  [hep-ph/9706529].
  
\bibitem{Kwiecinski:2001nh}
  J.~Kwiecinski, A.~D.~Martin, L.~Motyka and J.~Outhwaite,
  Phys.\ Lett.\ B {\bf 514} (2001) 355
  [hep-ph/0105039].

\bibitem{Vera:2006un}
  A.~Sabio Vera,
  Nucl.\ Phys.\ B {\bf 746} (2006) 1
  [hep-ph/0602250].
  
\bibitem{Vera:2007kn}
  A.~Sabio Vera and F.~Schwennsen,
  Nucl.\ Phys.\ B {\bf 776} (2007) 170
  [hep-ph/0702158 [HEP-PH]].

\bibitem{Angioni:2011wj} 
  M.~Angioni, G.~Chachamis, J.~D.~Madrigal and A.~Sabio Vera,
  Phys.\ Rev.\ Lett.\  {\bf 107}, 191601 (2011)
  doi:10.1103/PhysRevLett.107.191601
  [arXiv:1106.6172 [hep-th]].

\bibitem{Ducloue:2013wmi} 
  B.~Ducloue, L.~Szymanowski and S.~Wallon,
  JHEP {\bf 1305}, 096 (2013)
  [arXiv:1302.7012 [hep-ph]].

\bibitem{Colferai:2010wu} 
  D.~Colferai, F.~Schwennsen, L.~Szymanowski and S.~Wallon,
  JHEP {\bf 1012}, 026 (2010)
  [arXiv:1002.1365 [hep-ph]].


\bibitem{Ducloue:2013bva} 
  B.~Ducloue, L.~Szymanowski and S.~Wallon,
  Phys.\ Rev.\ Lett.\  {\bf 112}, 082003 (2014)
  [arXiv:1309.3229 [hep-ph]].

\bibitem{Caporale:2014gpa} 
  F.~Caporale, D.~Y.~Ivanov, B.~Murdaca and A.~Papa,
  Eur.\ Phys.\ J.\ C {\bf 74}, 3084 (2014)
  [arXiv:1407.8431 [hep-ph]].

\bibitem{Celiberto:2015yba} 
  F.~G.~Celiberto, D.~Y.~Ivanov, B.~Murdaca and A.~Papa,
  Eur.\ Phys.\ J.\ C {\bf 75}, no. 6, 292 (2015)
  doi:10.1140/epjc/s10052-015-3522-6
  [arXiv:1504.08233 [hep-ph]].

\bibitem{Safronov:2015bva} 
  G.~Safronov [CMS Collaboration],
  AIP Conf.\ Proc.\  {\bf 1654}, 040003 (2015)
  [arXiv:1501.02332 [hep-ex]].


\bibitem{Caporale:2015vya} 
  F.~Caporale, G.~Chachamis, B.~Murdaca and A.~Sabio~Vera,
  arXiv:1508.07711 [hep-ph].
  
\bibitem{Caporale:2015int} 
  F.~Caporale, F.~G.~Celiberto, G.~Chachamis and A.~Sabio~Vera,
  arXiv:1512.03364 [hep-ph].
  
  
\bibitem{Chachamis:2013rca}
  G.~Chachamis and A.~Sabio Vera,
  PoS DIS {\bf 2013} (2013) 167
  [arXiv:1307.7750].
  
\bibitem{Caporale:2013bva}
  F.~Caporale, G.~Chachamis, J.~D.~Madrigal, B.~Murdaca and A.~Sabio Vera,
  Phys.\ Lett.\ B {\bf 724} (2013) 127
  [arXiv:1305.1474 [hep-th]].
  
\bibitem{Chachamis:2012qw}
  G.~Chachamis, A.~Sabio Vera and C.~Salas,
  Phys.\ Rev.\ D {\bf 87} (2013) 1,  016007
  [arXiv:1211.6332 [hep-ph]].
  
\bibitem{Chachamis:2012fk}
  G.~Chachamis and A.~Sabio Vera,
  Phys.\ Lett.\ B {\bf 717} (2012) 458
  [arXiv:1206.3140 [hep-th]].
  
\bibitem{Chachamis:2011nz}
  G.~Chachamis and A.~Sabio Vera,
  Phys.\ Lett.\ B {\bf 709} (2012) 301
  [arXiv:1112.4162 [hep-th]].
  

\bibitem{Chachamis:2011rw}
  G.~Chachamis, M.~Deak, A.~Sabio Vera and P.~Stephens,
  Nucl.\ Phys.\ B {\bf 849} (2011) 28
  [arXiv:1102.1890 [hep-ph]].

\bibitem{Chachamis:2015zzp} 
  G.~Chachamis and A.~Sabio Vera,
  arXiv:1511.03548 [hep-ph].

  
\bibitem{Chachamis:2015ico} 
  G.~Chachamis and A.~Sabio~Vera,
  arXiv:1512.03603 [hep-ph].



\end{thebibliography}
\end{document}